\documentclass[journal]{vgtc}                %

\ifpdf%
  \pdfoutput=1\relax                   %
  \usepackage{graphicx}                %
  \DeclareGraphicsExtensions{.pdf,.png,.jpg,.jpeg} %
\else%
  \usepackage{graphicx}                %
  \DeclareGraphicsExtensions{.eps}     %
\fi%

\graphicspath{{figures/}{pictures/}{images/}{./}} %

\usepackage{microtype}                 %
\PassOptionsToPackage{warn}{textcomp}  %
\usepackage{textcomp}                  %
\usepackage{mathptmx}                  %
\usepackage{times}                     %
\usepackage{cite}                      %
\usepackage{tabu}                      %
\usepackage{booktabs}                  %

\usepackage{enumitem}
\setlist{nolistsep}
\usepackage{subfigure}
\usepackage{setspace}

\usepackage{textcomp}
\usepackage{color,soul}
\usepackage{url}
\usepackage[colorinlistoftodos]{todonotes}
\usepackage{ifthen}
\usepackage{setspace}
\usepackage{enumitem}

\reversemarginpar %

\newcommand{\remark}[4]{%
\ifthenelse{\equal{#4}{m}}%
{%
\todo[caption=xxx,color=#3,size=\small]{%
\begin{spacing}{0.9}%
\textsf{\textbf{#1:}\\#2}%
\end{spacing}%
}%
}{
\begin{spacing}{1.2}
\todo[inline, color=#3,size=\small,caption=xxx]{
\textsf{\textbf{#1:}~#2}
}
\end{spacing}
}
}

\definecolor{remarkVahan}{HTML}{B3E2CD}
\definecolor{remarkTim}{HTML}{FDCDAC}
\definecolor{remarkKim}{HTML}{CBD5E8}
\definecolor{remarkMichael}{HTML}{F4CAE4}
\definecolor{remarkYalong}{HTML}{CCCCCC}

\definecolor{remarkLang}{HTML}{FB9A99}
\definecolor{remarkLangQ}{HTML}{FDB462}

\usepackage{caption}

\usepackage[final, todonotes={disable}]{changes}

\onlineid{0}

\vgtccategory{Research}
\vgtcpapertype{evaluation}

\title{Scalability of Network Visualisation \\ from a Cognitive Load Perspective}

\author{Vahan Yoghourdjian, Yalong Yang, Tim Dwyer, Lee Lawrence, Michael Wybrow and Kim Marriott}
\authorfooter{
\item
 Vahan Yoghourdjian, Tim Dwyer,  Michael Wybrow and Kim Marriott are with the Department of Human-Centred Computing, Faculty of Information Technology, Monash University, Melbourne, Australia. E-mail: \{vahan.yoghourdjian,tim.dwyer,michael.wybrow,kim.marriott\}@monash.edu
\item 
 Lee Lawrence is with the Faculty of Business and Economics, Monash University, Melbourne, Australia. E-mail: lee.lawrence@monash.edu
\item 
 Yalong Yang was with the Department of Human-Centred Computing, Faculty of Information Technology, Monash University, Melbourne, Australia. He is now with the School of Engineering and Applied Sciences, Harvard University, MA, USA. E-mail: yalongyang@g.harvard.edu
}

\shortauthortitle{Biv \MakeLowercase{\textit{et al.}}: Global Illumination for Fun and Profit}

\abstract{
Node-link diagrams are widely used to visualise networks. However, even the best network layout algorithms ultimately result in `hairball' visualisations when the graph  reaches a certain degree of complexity, requiring simplification through aggregation or interaction (such as filtering) to remain usable. Until now, there has been little data to indicate at what level of complexity node-link diagrams become ineffective or how visual complexity affects cognitive load. To this end, we conducted a controlled study to understand workload limits  for a task that requires a detailed understanding of the network topology---finding the shortest path between two nodes. We tested performance on graphs with 25 to 175 nodes with varying density. We collected performance measures (accuracy and response time), subjective feedback, and physiological measures (EEG, pupil dilation, and heart rate variability). To the best of our knowledge this is the first network visualisation study to include physiological measures. Our results show that people have significant  difficulty finding the shortest path  in high density  node-link diagrams with more than 50 nodes and even low density graphs with more than 100 nodes. From our collected EEG data we observe functional differences in brain activity between hard and easy tasks. We found that cognitive load increased up to certain level of difficulty after which it decreased, likely because participants had given up. We also explored the effects of global network layout features such as size or number of crossings, and features of the shortest path such as length or straightness on task difficulty. We found that global features generally had a greater impact than those of the shortest path.

} %


\CCScatlist{ %
 \CCScat{K.6.1}{Management of Computing and Information Systems}%
{Project and People Management}{Life Cycle};
 \CCScat{K.7.m}{The Computing Profession}{Miscellaneous}{Ethics}
}

\teaser{
  \centering
  \includegraphics[width=0.32\textwidth]{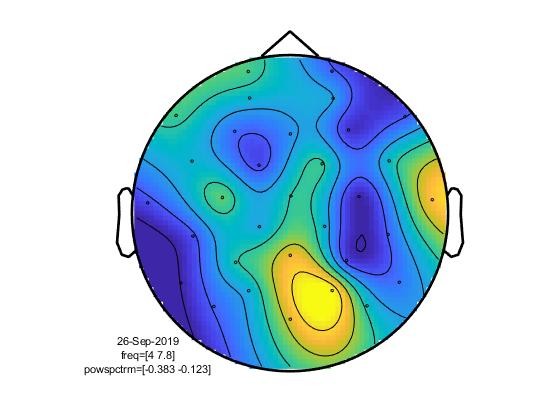}
  \includegraphics[width=0.32\textwidth]{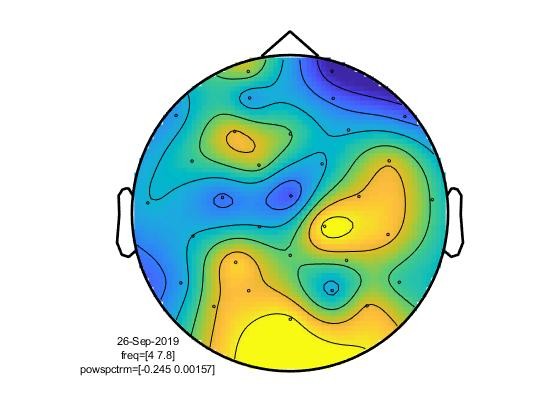}
  \includegraphics[width=0.32\textwidth]{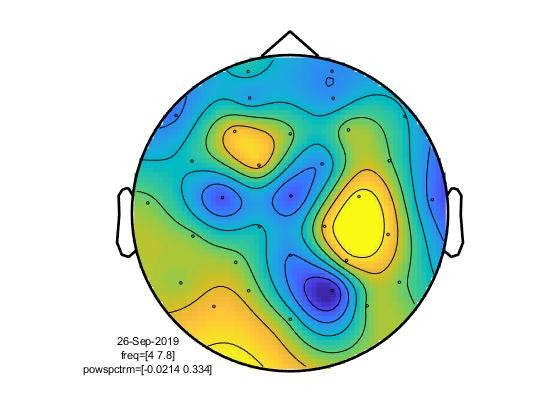}
  \caption{EEG topographical maps of median theta brain activity for (from left-to-right) easy tasks, hard tasks and the difference (hard-easy) for participants undertaking path-finding exercises on network diagrams.}
  \label{fig:eeg-topographic}
  \vspace{1em}
}

\vgtcinsertpkg

\begin{document}

\firstsection{Introduction\label{sec:introduction}} 
\maketitle
Visualisation helps analysts to understand and explain complex data. However, there exist factors that limit the amount of information that can be visualised. Scaling to a large number of data elements is a major issue in visualisation design.  Eick and Karr~\cite{eick2002visual} discuss how human perception, monitor resolution, visual metaphors, interactivity, data structures and algorithms, as well as computational infrastructure affect visual scalability. 
For network visualisation, the last five factors have been well explored~\cite{jankun2014scalability}. However, scalability of human perception remains under-studied. A recent survey of 152 experimental studies of node-link visualisation techniques found that most of the networks considered in these studies were relatively small and sparse \cite{oursurvey}.  The survey authors called for studies that control for the size and complexity of the network to explicitly test perceptual scalability of  network visualisation techniques. 

Here we address perceptual scalability of node-link diagrams, which are undoubtedly the most common way of visualising networks.  Surveys like that of Jankun~\textit{et al.}~\cite{jankun2014scalability} speak about the so-called `hair-ball effect', wherein, node-link diagram representations of larger small-world or scale-free graphs are no longer useful for understanding the connectivity of all but peripheral nodes in the visualisation. Previous studies suggest that, while matrix-based representations are more effective than node-link diagrams  for some tasks~\cite{ghoniem2004comparison,keller2006matrices,okoe2018node}, node-link diagrams are superior for connectivity tasks. For this reason we focus on the scalability of node-link diagrams for a widely used connectivity task, that of finding the shortest-path between two nodes.

We conducted an experiment in which 22 participants found the length of the shortest path between two nodes on 42 small-world networks, ranging from 25 to 175 nodes with three levels of density.  
In addition to task completion speed, accuracy and self-reported difficulty, we also collected physiological measures known to be associated with mental effort: brain electrical activity (EEG), heart rate, and pupil size. This is the first study that we know of to \replaced{provide a holistic analysis across subjective, physiological as well as performance measurements for a network visualisation task}{collect physiological data for a network visualisation task and one of the first to collect data for any visualisation task}.  Our main contributions are as follows.

\begin{enumerate} 
    \item We establish that the usefulness of node-link diagrams for finding shortest paths quickly deteriorates as the number of nodes and edges increases---as discussed in Section \ref{sec:scalability}. For small-world graphs with 50 or more nodes and a density (ratio of edges to nodes) of 6, participants were unable to correctly answer in more than half of the trials.  This was also the case for graphs with a density of 2 and more than 100 nodes.
    \item We provide an analysis of the relationship between \emph{task hardness} and the physiological measures of cognitive load---see Section \ref{sec:physiological}. We found that these measures of load increased with task hardness until a threshold is reached, after which it decreases, suggesting that participants give up. This analysis relied on combining accuracy and self-reported difficulty to give a single measure of task hardness for each of the 42 stimuli. 
    \item We make \replaced{an initial}{the first} identification of brain regions associated with a network visualisation task---in Section \ref{sec:eeganalysis}---and reveal functional differences in  brain activity between hard and easy tasks. \added{Here we used self-reported difficulty.}
    \item We explored the effects of global network layout features (such as size or number of crossings) and features of the shortest path (such as length or straightness) on task difficulty. We found that global features generally had a greater impact \added{on hardness} than those of the shortest path---see  Section \ref{sec:featureshardness}.   %
    \item Furthermore, measuring cognitive load through physiological measures requires careful setup and analysis. We believe that our experience and methodology will inform future visualisation researchers who also wish to use such measures to evaluate cognitive load for other kinds of visualisation tasks---Section \ref{sec:limit}.
\end{enumerate}

Our research adds to the understanding of the perceptual scalability of node-link diagrams. It informs visualisation designers about the size of network for which node-link diagrams are appropriate and at what point the number of nodes and links displayed to the user should be limited, e.g.,\  through  filtering or aggregation techniques. It also helps to clarify the visual features that layout algorithms should focus on to improve usability.

\section{Background and Related Work}
\label{sec:relwork}

\subsection{Network Visualisation Effectiveness Studies}

Task performance---accuracy and/or response time---is the standard measure of visualisation efficiency 
used across many network visualisation studies.
There has been a lot of research exploring the effects of layout features of node-link diagrams on task effectiveness,  many of which use shortest-path finding as a task \cite{purchase1995validating, purchase1997aesthetic, huang2008effects, huang2007using, kobourov2014crossings, dawson2015search}.  
In particular, a study by Ware~\emph{et al.}~\cite{ware2002cognitive} explores the effects of different layout features on response time in a shortest-path finding task on node-link diagrams, which they attribute to `cognitive cost'.  Their results indicate that the number of hops on the shortest path is the highest contributor to cognitive cost, followed by the straightness of the shortest path.

A number of empirical studies compare node-link diagrams with alternate visualisation types and techniques. For example, Ghoniem~\emph{et al.}~\cite{ghoniem2004comparison} compare the effectiveness of node-link diagrams with adjacency matrices for various tasks.  The study is unusual in testing relatively dense graphs, e.g.,\ up to 100 nodes and 3,600 edges.
For such graphs they found matrices provided better support than node-link diagrams for many tasks, the exception being path finding, which remains very difficult in matrices regardless of density.

A recent survey of network visualisation user studies has  explored the literature in terms of number of nodes and edges used in published studies~\cite{oursurvey}.  While there are many  studies evaluating different representations of network data, these rarely significantly vary the size of the data, preferring one or two data sets, carefully chosen  to be well within the capabilities of at least one of the techniques being tested (e.g.,\ \cite{lee2016communities, saket2015map, greffard2011visual}).  We are aware of a few studies that involve large graphs (e.g.,\ hundreds or thousands of nodes) \cite{moscovich2009topology, nekrasovski2006evaluation, marner2014gion, ware2005supporting, dunne2013motif, archambault2010readability}, but they all use interactive query or aggregation techniques, allowing the user to filter the input graph, so that only a small subset of the nodes and links are actually shown to the participant.

There is, therefore, a lack of evidence regarding the effectiveness of large node-link or other network visualisations for tasks that require a detailed understanding of the network structure. 
This work aims to determine the thresholds for node-link visualisations after which designers should limit the number of nodes and edges on display or switch to a summary representation~\cite{yoghourdjian2018graph}  in order to best cater for human perceptual and cognitive capabilities.

\subsection{Cognitive Load}

Cognitive Load Theory suggests that humans process information using  limited working memory~\cite{sweller1988cognitive}. The theory was initially developed in the fields of education and instructional design. %
 It distinguishes between \emph{intrinsic}, \emph{extraneous}, and \emph{germane} cognitive load. \emph{Intrinsic} cognitive load is associated with the inherent difficulty of the instruction or task. \emph{Extraneous} cognitive load depends on how the instruction and information is presented, while \emph{germane} cognitive load refers to processing, acquiring and automating schemata~\cite{chandler1991cognitive, sweller1998cognitive}.
Three main types of measures can be used to assess cognitive load: subjective feedback, performance-based (accuracy and response time), and physiological~\cite{de1996measurement} such as brain activity, pupil dilation or heart rate variability.

Only one study (to our knowledge) has directly used cognitive load as a measure to evaluate network visualisations. Huang~\emph{et al.}~\cite{huang2009measuring} conducted a study that explored cognitive load in node-link diagrams. They utilised an efficiency measure based on the approach proposed by Paas and Van Merrienboer~\cite{paas1993efficiency} for comparing instructional materials which combines self-reported mental effort and performance measures. Huang~\emph{et al.} manipulated complexities of the visual representation, data, and task, to show that cognitive load is affected by these factors.  Their results confirmed that the participant feedback matched their expectations in terms of task difficulty. However, the graphs they used were fairly small and they did not consider physiological measures of cognitive load.

Even though physiological measures have been frequently used to measure cognitive load in systems engineering and psychology, to our knowledge, there have been very few studies that use physiological measures to evaluate data visualisations and no studies that utilise them to measure the effectiveness of network visualisations. Instead researchers have almost totally relied on performance-based and subjective measures.

An exception is Anderson~\emph{et al.}~\cite{anderson2011user} who compared the cognitive load of participants when identifying the larger interquartile range on a variation of box plot types. They measured task difficulty, response time and brain activity using EEG. They used the spectral differences in the alpha and theta frequency bands of the signals acquired by EEG as an indicator of cognitive load. The results showed a correlation between these three measures, with an increase in response time and cognitive load as tasks became more difficult.

In another study Peck~\emph{et al.}~\cite{peck2013using} used functional near-infrared
spectroscopy to compare the cognitive load imposed by pie charts versus bar charts. They also used accuracy, response time, and subjective feedback (NASA-TLX). They asked participants to estimate differences between two highlighted sections, given either a pie or bar chart. The results did not show a difference in cognitive load between the two visualisation idioms.
This is perhaps attributable to the task not really involving problem solving, but relying mostly on visual perception.

\replaced{One}{Part of the} contribution of this paper is our initial exploration of the applicability of different physiological measures to network visualisation.

\subsection{EEG Measurement of Cognitive Load}
\begin{figure}
\centering
\includegraphics[width=0.85\columnwidth]{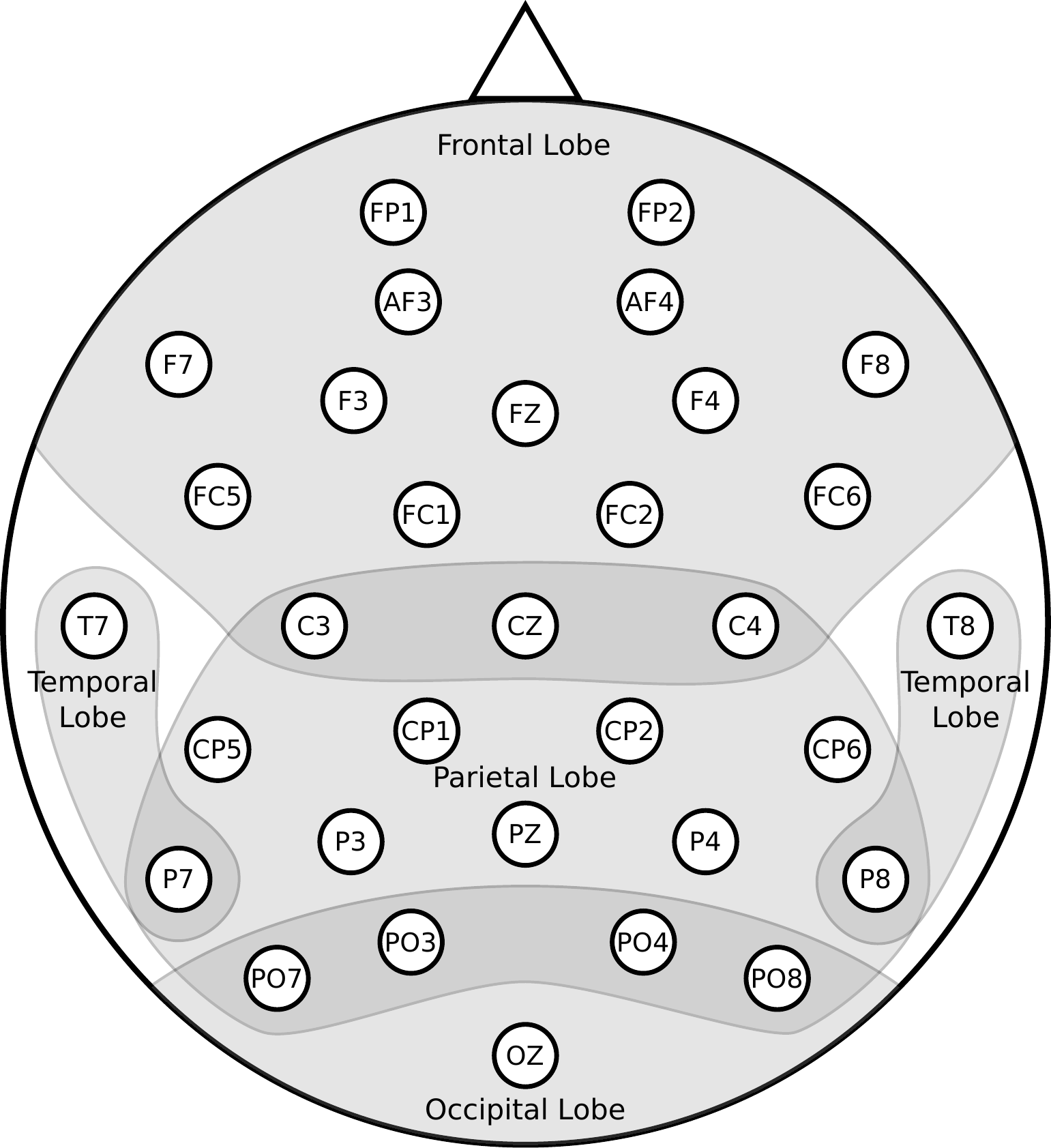}
\caption{The arrangement of the 32 dry electrodes of the \emph{g.Nautilus} EEG cap~\cite{gtec} used in our study, and a broad indication of the major brain regions they are associated with.  The placement is based on the International 10-20  system with extra electrodes. }
\label{fig:10-20}
\end{figure}

Quantitative EEG is the broad name given to the analysis of brain electrical activity with respect to its oscillations, or frequency components. Data is collected from electrodes placed in a standard configuration on the surface of the head---see Figure~\ref{fig:10-20}. Generally speaking, brain electrical oscillations occurring between 4 and 8 Hertz, called theta activity, have  been associated with memory processing, such as during memory encoding~\cite{klimesch_eeg_1999}, recognition or processing during spatial navigation~\cite{white2012brain}  and other related processes including error detection~\cite{trujillo_theta_2007}. 

Theta activity is also commonly used to measure  cognitive load. Increased activity is associated with increased cognitive load processing and task difficulty, typically at the central frontal lobe electrodes, FZ~\cite{dan_real_2017}. However, the region of the brain associated with increased activity depends upon the kind of task. For instance, a linguistic (i.e., hypertext) based task highlighted electrodes F7 and P3 as being more important~\cite{antonenko2010using}.

Briefly, whilst brain regions do not typically work in isolation, different brain areas have different roles.
Figure~\ref{fig:10-20} includes a schematic map of the major regions. Broadly speaking, the frontal lobe is involved in reward and error processing, impulse control, decision making, problem solving and abstract reasoning, motivation, language production, and motor planning, control and execution. The parietal lobe is generally involved in touch sensation, as well as visuo-spatial processing and perception, including mental imagery. The temporal lobe is involved in auditory sensation, object recognition, memory, language comprehension, and emotions. Finally, at the rear of the brain, the occipital cortex is involved in lower-level visual processing.

As no previous study has investigated brain activity for a network visualisation task,  it is difficult to predict precisely which areas of the brain and hence electrodes will be involved. For instance, finding the shortest path could involve memory encoding, decision making (including error detection), and spatial processing, within what  is a predominately spatial navigation decision task. 
A spatial navigation task in the literature implicated right temporal and bilateral parieto-occipital theta increases, with left posterior decreases~\cite{white2012brain}. However, this task is a less than ideal comparison because it did not manipulate cognitive load.

The brain imaging study with the most similar task did not use EEG analysis. Instead, it used  functional Magnetic Resonance Imaging (fMRI) to detect oxygen/blood flow in the brain as a measure of activity. Kaplan and colleagues~\cite{kaplan2017neural} sought to identify regions of brain activity during a maze processing task. Participants were required to decide what was the shortest path between their starting point and an end location. Some mazes only had one choice point,  while others had two choice points when deciding the shortest path. 
Their results suggested that there were brain activity differences depending on the number of decisions.
These results suggest that shortest path tasks could involve theta activity in the left-frontal, right-parietal, and left-temporal regions.

\section{User Study}
\label{sec:study}

Our study was designed to investigate the perceptual scalability of node-link diagrams for  graph connectivity tasks, identifying the graph complexity and size beyond which they cease to be useful for such tasks. This extends previous studies such as Ware~\cite{ware2002cognitive}
by considering a much greater range of graph sizes and densities.
We also explore physiological measures of cognitive load: EEG, pupil dilation and heart rate variability.

\subsection{Setup}
\noindent\textbf{Participants.}
The study had 22 participants: 14 male, 8 female. 18 participants were in age range 20--30, while 4 were aged 35--45. All participants had a background in Computer Science. The participants were asked about their familiarity with node-link diagrams and the shortest path problem. 9 participants frequently encountered node-link diagrams and the shortest path problem, while the remaining 13 participants occasionally did.

\noindent\textbf{Task.}
We chose a network connectivity task because this is a common high-level task and node-link diagrams have been found to be particularly effective for this task~\cite{ghoniem2004comparison}. Specifically, participants were shown a range of graphs and instructed to identify the shortest path between two highlighted nodes and determine the number of nodes on this path, if they could.

\noindent\textbf{Graph corpus.}
The graphs shown to the participants varied in two dimensions; numbers of nodes and edge density.  
Of course, other aspects of the graph structure can be expected to affect task complexity. However, to keep our study under two hours we were forced to consider only a single type of graph structure.  We wanted our generated graphs to be similar to real-world graphs. We chose to use the \emph{Barab\'asi-Albert} model~\cite{barabasi1999emergence} as this is known to produce graphs with small-world characteristics. Such graphs are common in nature and are frequently studied, e.g., in cell-biology \cite{albert2005scale}, bibliography \cite{farkas2002networks} and  internet topology~\cite{faloutsos1999power}.

We generated our stimuli using code written in \emph{JavaScript} and based on the \emph{Barab\'asi-Albert}~\cite{barabasi1999emergence} model.  We preferred not to use standard generators, since most require parameters specifying the total number of nodes and the number of edges to added at each iteration. Instead, we wanted to specify the total number of nodes and edges.

The number of nodes in the generated graphs ranged from 25  to 175 nodes in increments of 25. We experimented with different nodes ranges, but our pilot studies showed that task accuracy was at a maximum at 25 nodes, while the task was too difficult at 175 nodes.
To calculate the edges, we used densities of 2, 4 and 6, where $density = \mathit{number~of~edges} / \mathit{number~of~nodes}$. We chose these densities because real-world examples often have densities of less than 10 \cite{melancon2006just} and the results of our pilot studies showed that the graphs became unreadable beyond these values. We generated two graphs for each combination of number of nodes and density, giving 42 graphs, plus 3 graphs for training.
The graphs were arranged using the force-directed layout of \emph{WebCola}~\cite{webcola} and saved as drawings in SVG format.

For each graph we computed a start and end node of the path. We first selected the furthest node from the centre of the diagram and then selected the nearest node to the opposite side of the vector, passing through the centre. Due to the small-world nature of the graphs and the force-directed  layout, this  led to non-trivial shortest paths.

\begin{figure}
\centering
\includegraphics[width=0.85\columnwidth]{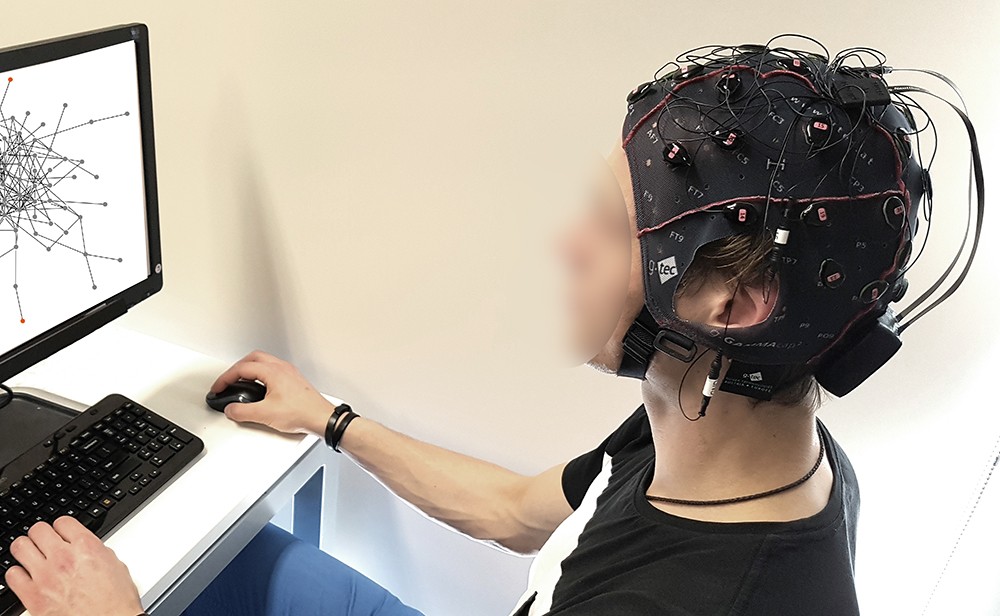}
 
\caption{One of the participants during the study (face obscured for anonymity) wearing the \emph{g.Nautilus} EEG cap. The eye-tracker device is mounted under the display and the heart rate monitor is worn under clothing.\label{fig:setup}}
\end{figure}
\noindent\textbf{Equipment.}	
\added{The study was conducted in a quiet office at Monash University with no natural light variation.}
The \deleted{study} set up is depicted in Figure~\ref{fig:setup}.
The study was run on a Windows 10 Dell Latitude E7440 laptop, equipped with 2.7 GHz i7 processor and 8 GB RAM. The visual representations were displayed in a $1920 \times 1080$ pixel area on a 22-inch HP monitor.  Mozilla Firefox 46.0 was used to to display the visualisations and collect participant responses.

A Tobii Pro X3-120 eye tracker~\cite{tobii} was used throughout the study. This was directly linked to the laptop.

A Polar H10 heart rate sensor was used to acquire heart rate information. This was linked to an iPhone~4 via Bluetooth and \emph{HRV Logger}~\cite{hrvlogger}.

A \emph{g.Nautilus}~\cite{gtec} electroencephalography (EEG) cap was also linked to the laptop to record the electrical activity of the brain via \emph{g.Recorder}; a software provided by \emph{g.tec}~\cite{gtec}. The cap exposes 32 data channels with dry electrodes spatially organised, based on the International 10-20 EEG placement system, with Modified Combinatorial Nomenclature as shown in Figure~\ref{fig:10-20}. 
Additional reference and ground electrodes were attached to the back of the participants' ears.  The EEG sampling was set to 250 Hz. An analogue bandpass filter was applied between 0.5 Hz and 100 Hz. A notch filter was used to suppress 48 Hz to 52 Hz power line interference. Sensitivity was set to +/- 2250 mV.

\subsection{Procedure}
The participants were shown an explanatory statement and were asked to sign a consent form. They were then presented with a short tutorial explaining the concept of shortest path and the task requirements. 

For each experimental task, the start and end pair of nodes were highlighted in orange and participants were asked to find the shortest path, taking note of the number of nodes between these end nodes. The correct answers for our tasks ranged from one to six. We also allowed participants to answer with `unsure' so that they did not need to guess. See Figure~\ref{fig:sp} for an example of the task. 
Both the answer and the time taken were recorded.

Each participant had to perform the task 45 times, of which 3 were training. The stimuli were shown in randomised order such that no consecutive graph contained the same number of nodes or number of edges. All graphs were shown to each participant using this order but starting with a different graph, resulting in an incomplete Latin Square design.  Each task was preceded by five seconds of blank screen to serve as a rest period, which also served as baseline for the physiological measures.

After each task, the participants   were asked to rate its difficulty. They were given a nine-grade symmetrical category scale used by Huang~\emph{et al.}~\cite{huang2009measuring} and evaluated by Bratfisch~\emph{et al.}~\cite{bratfisch1972perceived}. The scale uses the following terms: `very very easy', `very easy', `easy', `rather easy', `neither easy nor difficult', `rather difficult', `difficult', `very difficult', `very very difficult'.

\added{The participants were allowed to take breaks between each task. The breaks were not timed and the respective physiological measures were excluded from the analysis. Moreover, no fatigue was reported, or signs of fatigue observed.}

 Unlike Ghoniem~\emph{et al.}~\cite{ghoniem2004comparison} who allowed the participants to interact with the visualisation by highlighting neighbouring nodes when hovering over a specific node, we did not allow any interaction. We wanted to keep the variables of the study at a minimum in order to understand the basics of cognitive load and scalability. \replaced{Even though the participants had access to a mouse, they were asked not to use it during the task, and only used it to submit their answers.}{However, participants did have access to a mouse and some used the mouse cursor to follow potential paths in the graphs.}

\section{Performance and Subjective Measures}
\label{sec:scalability}
Our dependent measures fall into three categories: \emph{performance} (completion time and accuracy), \emph{subjective} (self-reported difficulty) and \emph{physiological} (pupil dilation, heart rate and EEG). The overall logic of our data analysis is 
\begin{enumerate}
	\item Determine scalability of the task in terms of performance and subjective measures.
	\item Develop a measure of task \textbf{\emph{hardness}} for each of the stimuli based on  performance and subjective measures. 
	\item Determine which regions of the brain are involved for the EEG analysis.
	\item Investigate the relationship between hardness and the physiological measures of cognitive load.
	\item Explain hardness in terms of \emph{graph metrics}.
\end{enumerate}
In this section we focus on the first two steps.

\begin{figure}
\centering
\includegraphics[width=\columnwidth]{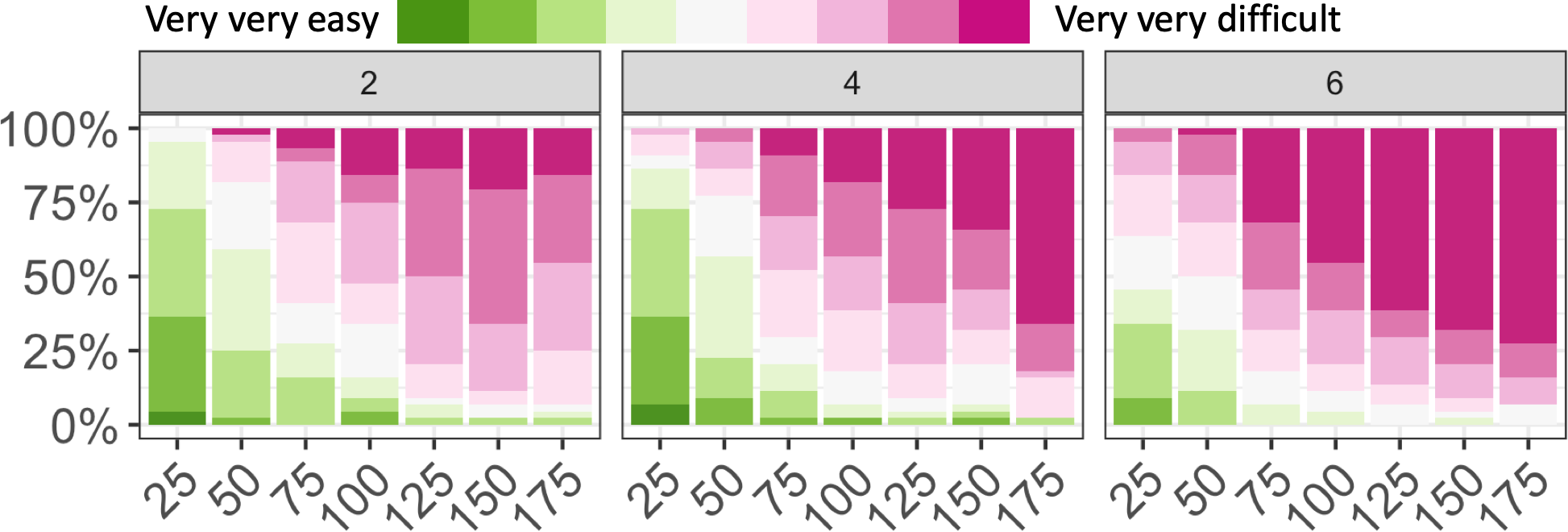}
\caption{Self-reported difficulty rating for the different sizes of graphs and densities. }
\label{fig:performance-difficulty}
\end{figure}

\begin{figure}
\centering
\includegraphics[width=\columnwidth]{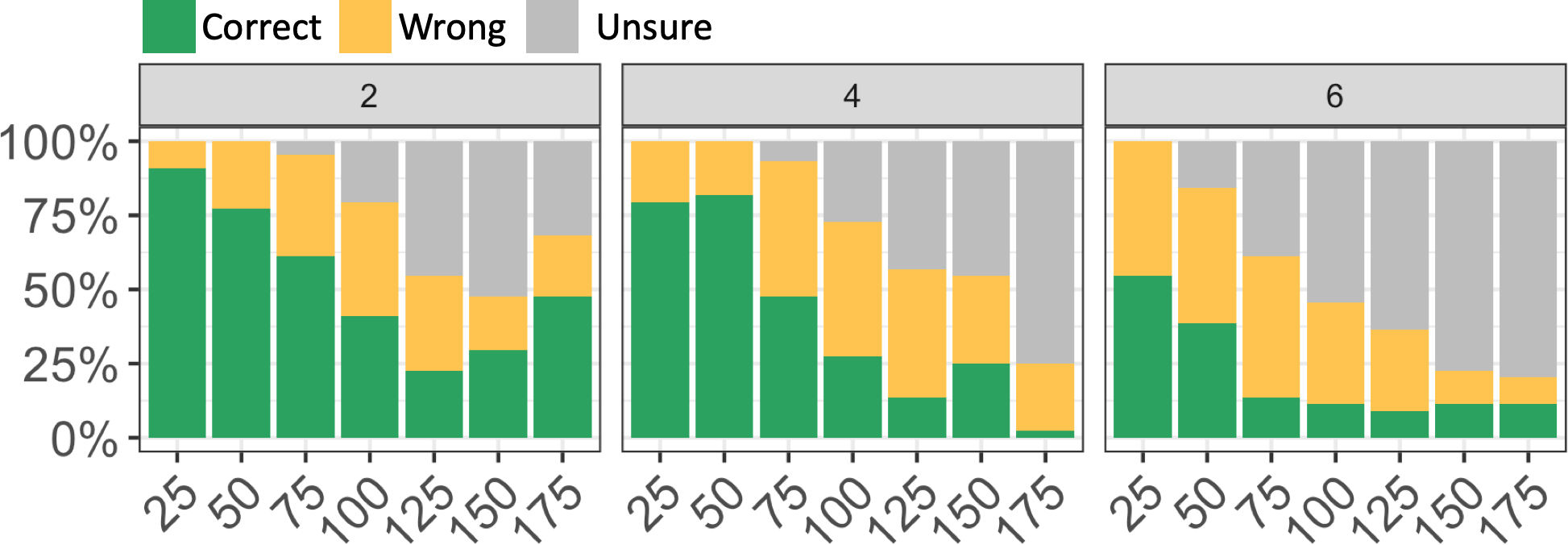}
\caption{Correctness for the different sizes of graphs and densities.}
\label{fig:performance-accuracy}
\end{figure}

\begin{figure}
\centering
\includegraphics[width=\columnwidth]{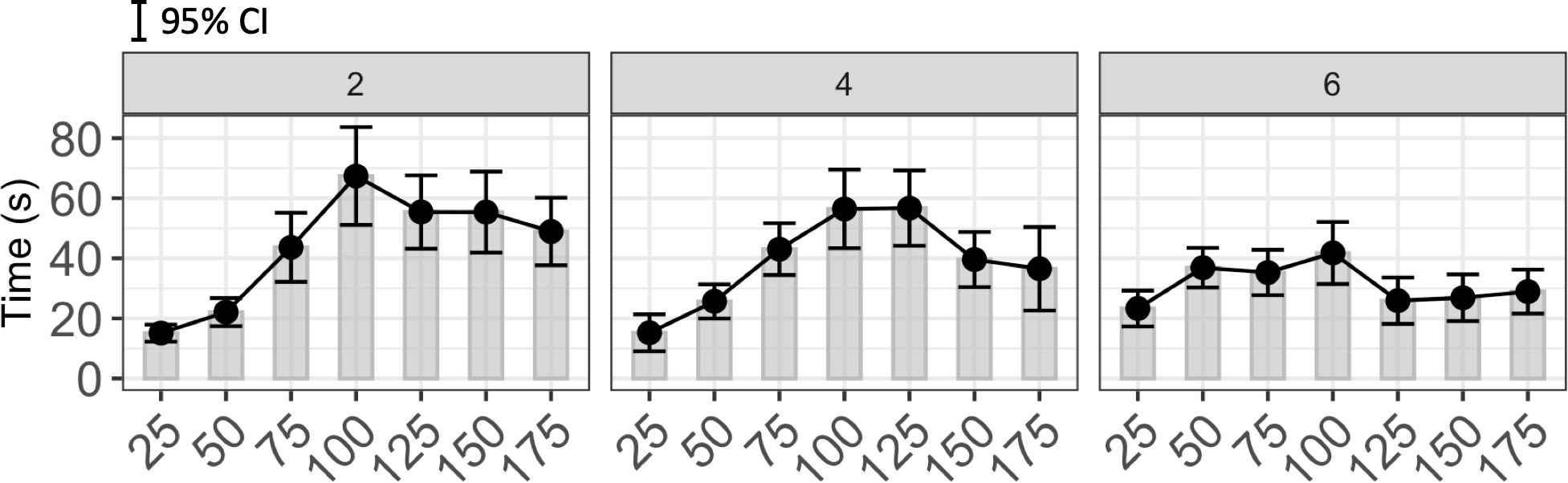}
\caption{Response time for the different sizes of graphs and densities.}
\label{fig:performance-time}
\end{figure}

\subsection{Scalability}
Figures~\ref{fig:performance-difficulty}, \ref{fig:performance-accuracy} and \ref{fig:performance-time} respectively show the self-reported difficulty, accuracy and response time of the participants in seconds with respect to each stimulus. Stimuli are grouped by density.

We used~\emph{repeated measures correlation}~\cite{bakdash2017} to investigate the correlation between  these measures and the number of nodes overall and for each density.

\noindent\textbf{Difficulty:} Overall, there is  a strong correlation between number of nodes and self-reported difficulty: \\
  $r_{rm}(901)=0.71$, 95\% CI $[0.67, 0.74]$, $p < 0.0001$.  \\ 
This strong correlation holds for each of the three densities:\\
Density 2: $r_{rm}(285)=0.76$, 95\% CI $[0.70, 0.80]$, $p < 0.0001$.\\
Density 4: $r_{rm}(285)=0.75$, 95\% CI $[0.70, 0.80]$, $p < 0.0001$.\\
Density 6: $r_{rm}(285)=0.72$, 95\% CI $[0.66, 0.77]$, $p < 0.0001$. 

\noindent\textbf{Accuracy:} 
\added{We weighted correct as 0, incorrect as 1. It is less clear how to weigh unsure. We cannot remove them entirely as failure to complete the task is an important signifier of task difficulty. One could argue that an unsure response indicates that the participant found it harder than those who gave an incorrect answer, as even with the wrong answer they at least felt that they could answer the question. We evaluated the effect on correlation (see supplementary materials). With a weight of 1 for unsure the correlation is 0.42. Correlation increases to 0.55 with a weight of 2 and then levels out. We therefore decided to weigh unsure as 2. Note that we redid the analysis for weighing unsure as 1 and it makes little difference (see supplementary materials).}\\
$r_{rm}(901)=0.55$, 95\% CI $[0.50, 0.59]$, $p < 0.0001$.\\
This strong correlation also holds for each of the three densities:\\
Density 2: $r_{rm}(285)=0.48$, 95\% CI $[0.39, 0.57]$, $p < 0.0001$.\\
Density 4: $r_{rm}(285)=0.68$, 95\% CI $[0.61, 0.74]$, $p < 0.0001$.\\
Density 6: $r_{rm}(285)=0.59$, 95\% CI $[0.51, 0.66]$, $p < 0.0001$.

\noindent\textbf{Time:} 
Overall, there is a correlation between number of nodes and response time:\\
$r_{rm}(901)=0.22$, 95\% CI $[0.15, 0.28]$, $p < 0.0001$.\\
However, while this correlation holds for density 2 and 4 it does not hold for density 6:\\
Density 2: $r_{rm}(285)=0.37$, 95\% CI $[0.27, 0.47]$, $p < 0.0001$.\\
Density 4: $r_{rm}(285)=0.24$, 95\% CI $[0.13, 0.35]$, $p < 0.0001$.\\
Density 6: $r_{rm}(285)=-0.04$, 95\% CI $[-0.16, 0.07]$, $p = 0.4592$. \\
One reason for this is might be that with the larger and denser examples participants  quickly realise that the task is too difficult and select `unsure' actually reducing their response time. For this reason we also checked the correlation when times for unsure responses were excluded. As expected this strengthens the correlation.  \\
Overall: $r_{rm}(601)=0.42$, 95\% CI $[0.35, 0.49]$, $p < 0.0001$.\\
Density 2: $r_{rm}(217)=0.48$, 95\% CI $[0.37, 0.57]$, $p < 0.0001$.\\
Density 4: $r_{rm}(198)=0.41$, 95\% CI $[0.29, 0.52]$, $p < 0.0001$.\\
Density 6: $r_{rm}(140)=0.31$, 95\% CI $[0.16, 0.46]$, $p = 0.0001$. \\

What is striking about these results is how hard participants find the task. 
Participants are: wrong or unsure more than 50\% of the time for graphs with 100 or more nodes for density 2;
they are wrong or unsure more than 50\% of the time for graphs with 75 or more nodes for density 4 and wrong or unsure more than 50\% of the time for graphs with 50 or more nodes for density 6. Indeed for density 6 and graphs with 75 or more nodes participant accuracy is around 16.67\%, the value we would expect by random selection. 

Our results strongly suggest that for path-based connectivity tasks node-link diagrams do not scale. Even for low-density graphs we find that determining shortest paths is only reasonable for graphs with less than 100 nodes and for higher-density graphs no more than 50 nodes.

\subsection{Task Hardness}
\label{sec:taskhardness}
For the subsequent analyses it is useful to have a single measure of the task hardness for each of the stimuli (graph and shortest path). Of course we don't have a direct measure of this but accuracy, response time and self-reported difficulty are all possible proxy measures with task hardness an underlying latent variable. 
There are a number of ways to do latent variable analysis. Basically the observed variables are modelled as linear combinations of the potential latent variables, plus ``error" terms. 
We used Principal Axis Factoring to extract the latent variable as it is one of the standard techniques used in psychological data analysis~\cite{costello_best_2019,kahn_factor_2006}.
Based on the prior discussion we chose not to use response times for unsure responses.

The analysis had three steps.
\begin{enumerate}
	\item We first normalised the measures for each participant in order to better take account of individual differences. The normalised score was simply the $z$-score of the measure w.r.t.\ all responses of the participant. 
	\item For each normalised measure we calculated the mean score for all participants for each of the 42 items (questions). 
	\item We then conducted Principal Axis Factoring, with the first principal component giving an estimate of \emph{task hardness}. 
	This indicated that 78\% of the variance in responses can be explained by task hardness and that the factor loadings for difficulty, accuracy and time were 1.04, 0.87 and 0.71 respectively.
\end{enumerate}
 
\section{Physiological Measures of Cognitive Load}
\label{sec:physiological}

In the next part of our analysis we explored the relationship between task difficulty and the physiological measures of cognitive load.  

\subsection{Data Preprocessing}

\noindent\textbf{Pupil Dilation:} We used Tobii Studio to record the eye tracker data from the Tobii Pro X3-120. We used the average of the two eyes in order to reduce noise. In cases where we had pupil size information for just one eye, we used that alone. 
For each task, we used the five seconds pre-task resting period to extract an average baseline, then we calculated pupil dilation by subtracting the average pupil size during the inter-trial rest period from the peak pupil size during task performance. We used peak dilation instead of mean pupil dilation since the latter does not work well with tasks that vary in length across participants \cite{beatty2000pupillary}. 
We used z-score to normalise the pupil dilation for each participant.

\noindent\textbf{Heart Rate Variability:} 
The polar H10 heart rate monitor recorded the beats per minute (bpm) and $r$-$r$ interval for each participant.
We used root mean squared successive difference (RMSSD)~\cite{haapalainen_psycho-physiological_2010,shaffer_overview_2017} which is a common measure for heart rate variability analysis~\cite{rowe_heart_1998} and used $z$-score to normalise this for each participant.

\noindent\textbf{EEG:} 
G.tec's \emph{g.Nautilis} dry 32 channel EEG system was used to record and digitise EEG using g.Recorder. Online, left ground and right reference ears were used in accordance to technical recommendations~\cite{gNautilusInstructions}. Raw EEG was converted to a format compatible with and then analysed using Fieldtrip~\cite{oostenveld_fieldtrip_2011} within MATLAB. Offline pre-processing consisted of re-referencing to the electrode average. Afterwards, the data was first visualised so that bad electrodes could be identified and interpolated using symmetrically chosen electrodes within a 5 cm radius. EEG data pre-processing continued after using a 1 to 30 hertz FIR band pass filter on whole data, using a hamming, 53dB/octave slope. These filters allowed a reduction of slow wave potentials whilst keeping the traditional shape of the eye blink response. Otherwise, this range was chosen to attenuate a level of noise associated with signals outside the range of frequency interest for this study whilst maintaining the ability to visualise muscle activity for later epoch rejection. After this, PCA ocular correction algorithms were performed on whole data to remove blinks and eye movements from the EEG data. 

Each participant's data was epoched for 5 seconds before the point where a participant signalled they had an answer. This was decided because there was a large variation in individual response times, with some trials taking over 2 minutes. This variability meant there was no guarantee that participants were concentrating for the entire time. We felt that epoching the 5 seconds prior to indicating an answer \deleted{had been reached} meant that the EEG results would be more comparable across trials and participants, as, at this point in time, they were more likely to be fully engaged in the task \added{(see supplementary materials for further discussion)}. 

We chose to analyse theta frequencies as theta power has been more consistently found to increase with cognitive load than other frequencies such as alpha whose power has been found to both increase and decrease with cognitive load~\cite{castro2020validating}.   Therefore, after epoching, FFT was performed on each stimuli, exporting 4 to 7.8 hertz as absolute power, using intervals of 0.2 and a 1 Hz taper. 

The data was converted into $z$-scores to normalise between participants.  Despite the cleaning algorithms, EEG outliers were found in the data that seemingly related to the amplification of noise, which did not seem to have any particular pattern within and between participants. To overcome this, it was decided to use the non-parametric estimates (i.e., median rather than mean) in all analyses involving EEG data. 

Trouble was found with two participants' EEG data---one participant's recording dropped out during online recording, and the other had reference problems---leaving EEG data from twenty participants for analysis.

\subsection{EEG Analysis}
\label{sec:eeganalysis}

As discussed in Section~\ref{sec:relwork}, different regions of the brain are associated with different functions. As a first step in our analysis of the EEG data we wished to determine which regions were involved in the shortest path task.

For each participant we split the stimuli into easy and hard tasks based on the individual participant's subjective ranking. We used the individual subjective ranking rather than the task hardness as we felt that this would better reflect the difficulty that that individual found with the stimuli. Even if a stimuli was generally found hard it could be that some participants found it easy just because they were lucky and happened to quickly see the shortest path.
We then computed the median theta power for the easy and hard stimuli at the different electrode locations, giving the EEG topographical maps shown in Figure~\ref{fig:eeg-topographic}. 

For the easy tasks the main activation is at the rear of the brain and slightly to the right in the parietal and occipital regions. There is also activation in the temporal region and little activity elsewhere. This pattern is similar to that previously found for spatial navigation~\cite{white2012brain}. It suggests that the decision making for these tasks is essentially visuo-spatial and that during their final decision, participants mostly relied on perceptually estimating the length of all possible routes.

On the other hand, when we look at the activation for hard instances there is activity on both sides of the occipital and parietal cortex and the right parietal and frontal regions and the left frontal region.
This pattern of activation is more similar to that found in~\cite{kaplan2017neural}. 
It suggests that for these stimuli a much more systematic step-by-step process is being used to find the shortest path with participants keeping track of the best path found so far in memory for comparison with the path under consideration. 

These distinctions are evident in the difference EEG topographical map. Increased activation in the left parieto-occipital region for the harder instances is conjectured to  reflect greater use of memory and pattern recognition (i.e., comparing the memory trace of the current path to previously considered paths~\cite{jacobs_eeg_2006}, and/or possibly pattern recognition more broadly when considering the role of the posterior temporal lobe~\cite{gaser_brain_2003}. On the other hand, the centro-parietal activity on the right hand side could represent the specialisation of the right ventro-parietal cortex for non-language-based spatial tasks, reflecting attention processing that also includes memory processes~\cite{cabeza_cognitive_2012,kaplan_neural_2017}. Finally, the frontal region activity on the left side is probably explained by traditional cognitive load theory~\cite{antonenko2010using,dan_real_2017}  and reflects semantic encoding and possibly retrieval processing~\cite{cabeza_imaging_2000,werkle-bergner_cortical_2006}  and working memory~\cite{blumenfeld_putting_2011,cabeza_imaging_2000}.

This analysis and the difference map suggests that electrodes in the left frontal region (F3, FC1), right centro-parietal (C4, CP2, CP6) and left parieto-occipital (PO7, P7) are the most likely electrodes to indicate increased cognitive load for our task. That said, noteworthy trace activation was also found in the right frontal region (F4) but not as strong as the other electrodes previously mentioned. 
We used the Wilcoxon signed rank test to evaluate the effect of easy vs. hard.
\added{We also calculated the $r$ effect size of these tests. We can interpret the $r$ effect size using Cohen's classification of effect sizes which is 0.1 (small effect), 0.3 (moderate effect) and 0.5 and above (large effect)~\cite{cohen2013statistical}.}
We note that the differences for these electrodes between the hard and easy stimuli are statistically significant \added{and all have large effects:
F3 ($p=.0066,~r=.62$),
FC1 ($p=.0034,~r=.66$),
F4 ($p=.0182,~r=.55$),
C4 ($p=.0056,~r=.63$),
CP2 ($p=.0077,~r=.61$),
CP6 ($p=.0023,~r=.68$),
P4 ($p=.0056,~r=.63$),
PO7 ($p=.0001,~r=.81$),
P7 ($p=.0028,~r=.67$).}

\subsection{Correlation with Task Difficulty} 

We next used repeated  measures correlation to investigate the correlation between the physiological measures discussed above and the task hardness. \textbf{We did not include unsure responses}. For EEG data, we considered theta for the electrodes identified in the previous section (F3, FC1, F4, C4, CP2, CP6, P4, PO7, P7). 

Table~\ref{tab:phy-correlation} shows only pupil dilation and heart rate variability demonstrated  a statistically significant positive correlation with hardness. Even then the correlations were not strong based on their correlation coefficients. This lack of correlation was  surprising as we would have expected cognitive load to be highly correlated with task difficulty.  

\begin{figure}
    \vspace{0.5em}
	\centering
	\includegraphics[width=0.8\columnwidth]{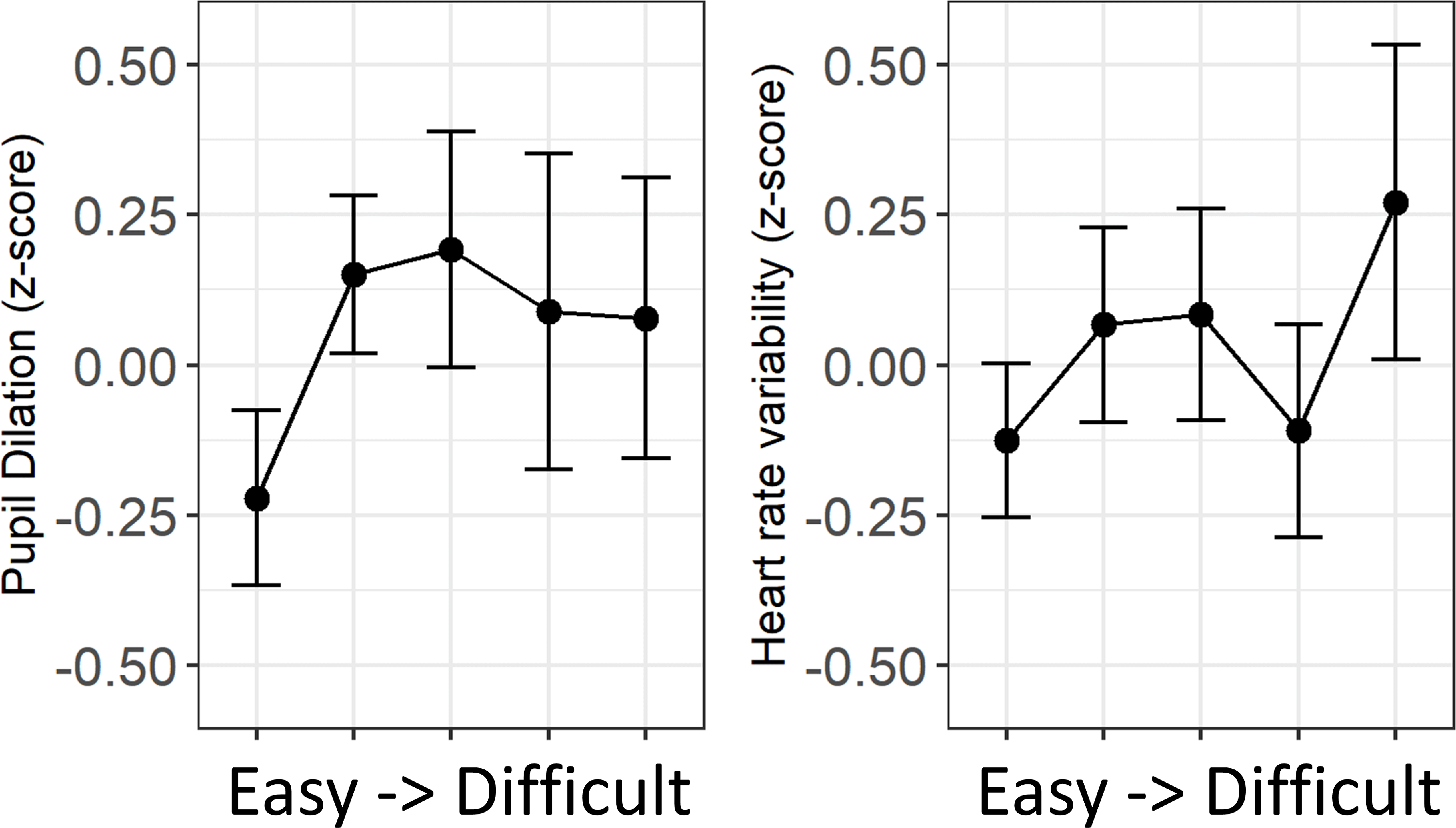}
	\caption{Pupil dilation and heart rate variability as a function of task hardness.\label{fig:eye-heart}}
\end{figure}

\begin{figure*}
	\centering
	\includegraphics[width=\textwidth]{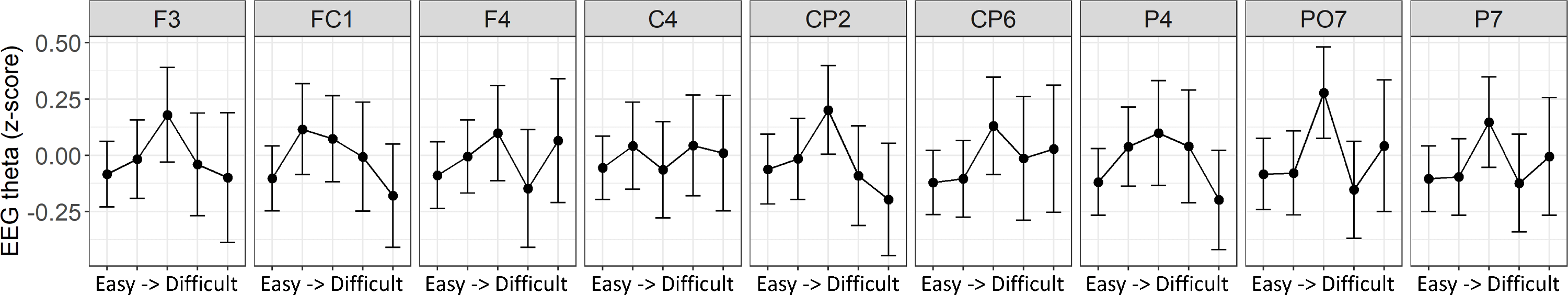}
	\caption{EEG measures as a function of task hardness.\label{fig:eeg}}
\end{figure*}

In order to better understand the relationship between the physiological measures and task hardness, we binned the measurements using a \emph{quantile interval} (i.e. each category contains an equal number of tasks) method to divide the range of hardnesss into five categories from easy to hard. We then plotted the 95\% CI of each measure as shown in Figures~\ref{fig:eye-heart} and~\ref{fig:eeg}.

We  see for pupil dilation and for most of the EEG measures (F3, FC1, CP2, CP6, P4, PO7, P7) that they first increase with task hardness but then decrease. This is why they exhibit only a weak correlation with task hardness. We conjecture that this is because once the task becomes very difficult participants switch off and no longer make the effort to find the right answer so cognitive load actually decreases. While we might have expected this for the unsure answers,  our results suggest that this happens even if the participants do not indicate they are unsure. For heart rate the story is not so straightforward but may be because heart rate is also influenced by stress and so participants' cognitive load decreased but stress was increased resulting in the overall increase.

\begin{table}
    \centering
    \includegraphics[width=\columnwidth]{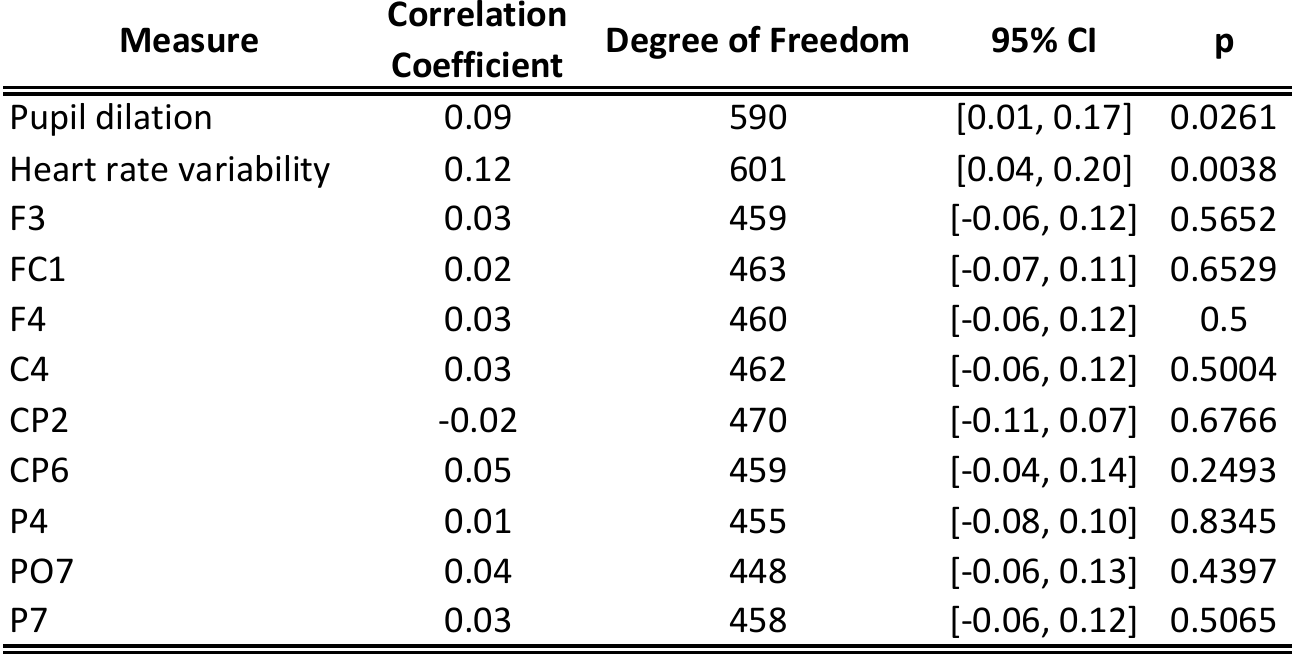}
    \caption{Correlation between physiological measures and hardness. 
    }
    \label{tab:phy-correlation}
\end{table}

\section{Graph and Layout Features Affecting Task Hardness}
\label{sec:featureshardness}

Clearly the size/complexity of the underlying graph affects the difficulty of finding the shortest path. As discussed in Section~\ref{sec:relwork} a number of papers have suggested other features that impact on this: length, i.e., number of nodes on the shortest path~\cite{ware2002cognitive}, number of crossings and crossing angle on the shortest path~\cite{ware2002cognitive,huang2008effects}, directness of shortest path~\cite{ware2002cognitive}, and degree of nodes on the shortest path~\cite{ware2002cognitive}. 
Given that we have a measure of task hardness we decided to conduct an  analysis investigating  how different features of the overall graph and of the shortest path(s) between the two target nodes affected task hardness. As our stimuli had not been designed to directly answer this question our analysis is necessarily limited but is still illuminating. 

\subsection{Graph and Layout Features}
One limitation of the stimuli is that most of them contained more than one shortest path. Therefore, for most features we provide both the total value over all of the shortest paths and the average value overall all shortest paths. We would expect the first value to be more highly correlated if participants examine all shortest paths but the second to be more highly correlated if they only examine one (or few) of the shortest paths.
We also consider a global measure of crossings, e.g. the total number of crossings. The rational for this is that other paths in the graph must be examined to complete the task, not just the shortest path itself. 
In some cases it is unclear how best to measure a particular feature so we used a number of metrics. The complete list of metrics is:

\noindent
Measures of size/complexity:
\begin{itemize}
	\item \emph{nodes}: number of nodes 
	\item \emph{density}: $nodes/edges$ 
	\item \emph{edges}: number of edges 
\end{itemize}
Measures of crossings and crossing angle:
\begin{itemize}
    \item \emph{gLLCrossingCount}: total number of link-link crossings 
	\item \emph{gLNCrossingCount}: total number of link-node crossings 
	\item \emph{gCrossingAngle}: overall sum of the angles at which links cross 
	\item \emph{gCrossingCount}: total number of link-link and link-node crossings
	\item \emph{gCrossingLLAngleLNCount}: the overall sum of the angles at which links cross + the number of link-node crossings
	\item \emph{sLLCrossingCount}: total number of link-link crossings on the shortest paths
	\item \emph{dsLLCrossingCount}:  average number of link-link crossings on the shortest paths
	\item \emph{sLNCrossingCount}: total number of node-link crossings on the shortest paths
	\item \emph{dsLNCrossingCount}: average number of node-link crossings on the shortest paths
	\item \emph{sCrossingAngle}: overall sum of the angles at which links cross on the shortest paths
	\item \emph{dsCrossingAngle}: average sum of the angles at which links cross on the shortest paths
\end{itemize}
Length of shortest path:
\begin{itemize}
	\item \emph{LengthOfShortestPath}: Number of nodes on the shortest path
	\item \emph{sEuclidean}: total Euclidean distance of shortest paths
	\item \emph{dsEuclidean}: average Euclidean distance of shortest paths
\end{itemize}
Degrees of nodes on the shortest path:
\begin{itemize}
	\item \emph{sDegrees}: total sum of the degrees of nodes on the shortest paths
	\item \emph{dsDegrees}: average total sum of the degrees of nodes on the shortest paths 
\end{itemize}
Straightness of shortest path:
\begin{itemize}
	\item \emph{sEquator}: total geodesic path deviation on the shortest paths
	\item \emph{dsEquator}: avg.\ total geodesic path deviation on shortest paths 
	\item \emph{sTurningAngle}: sum of turning angles on the shortest paths
	\item \emph{dsTurningAngle}: avg.\ sum of turning angles on shortest paths
\end{itemize}

The penalty for small angles was calculated by measuring the angle between two link crossings and subtracting it from 90 degrees. This applies a high penalty for small angles, but a small penalty for crossings where the links are close to orthogonal.

\begin{figure*}
    \centering
    \subfigure[Shortest paths]{\includegraphics[width=0.3\linewidth]{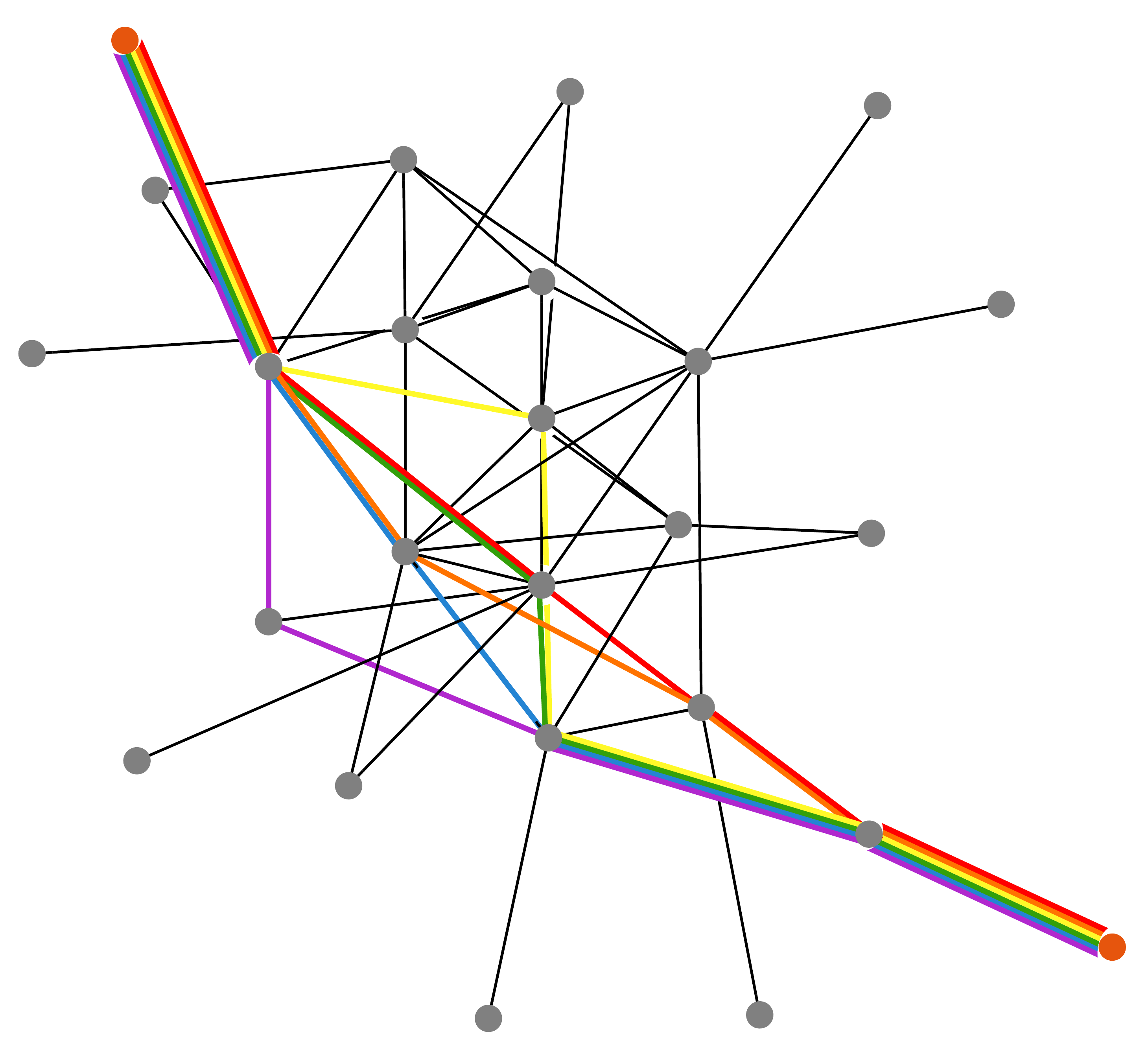}
    \label{fig:sp}}
    \subfigure[Link-link crossings]{\includegraphics[width=0.3\linewidth]{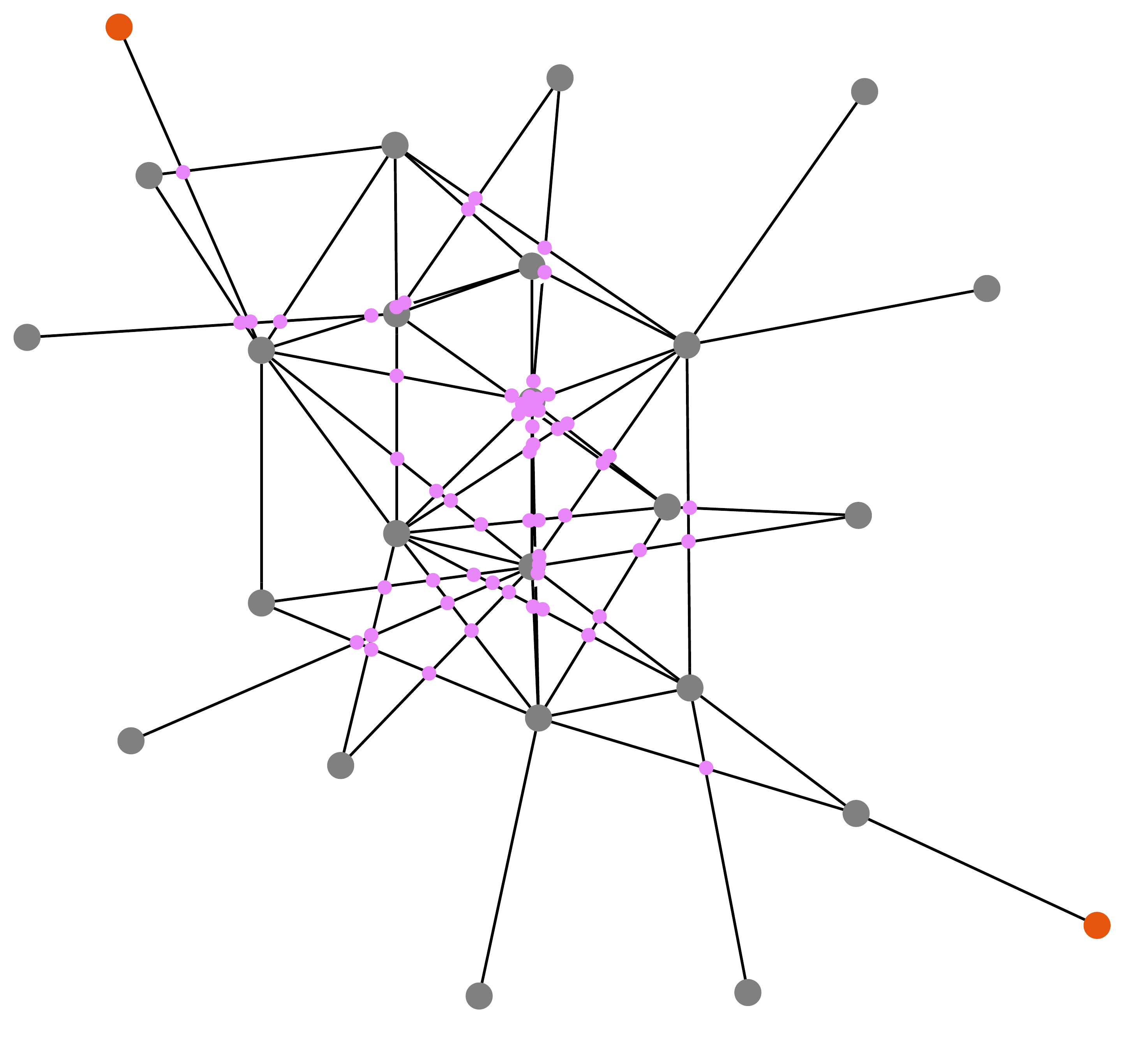}\label{fig:gLL}}
    \subfigure[Node-link crossings]{\includegraphics[width=0.3\linewidth]{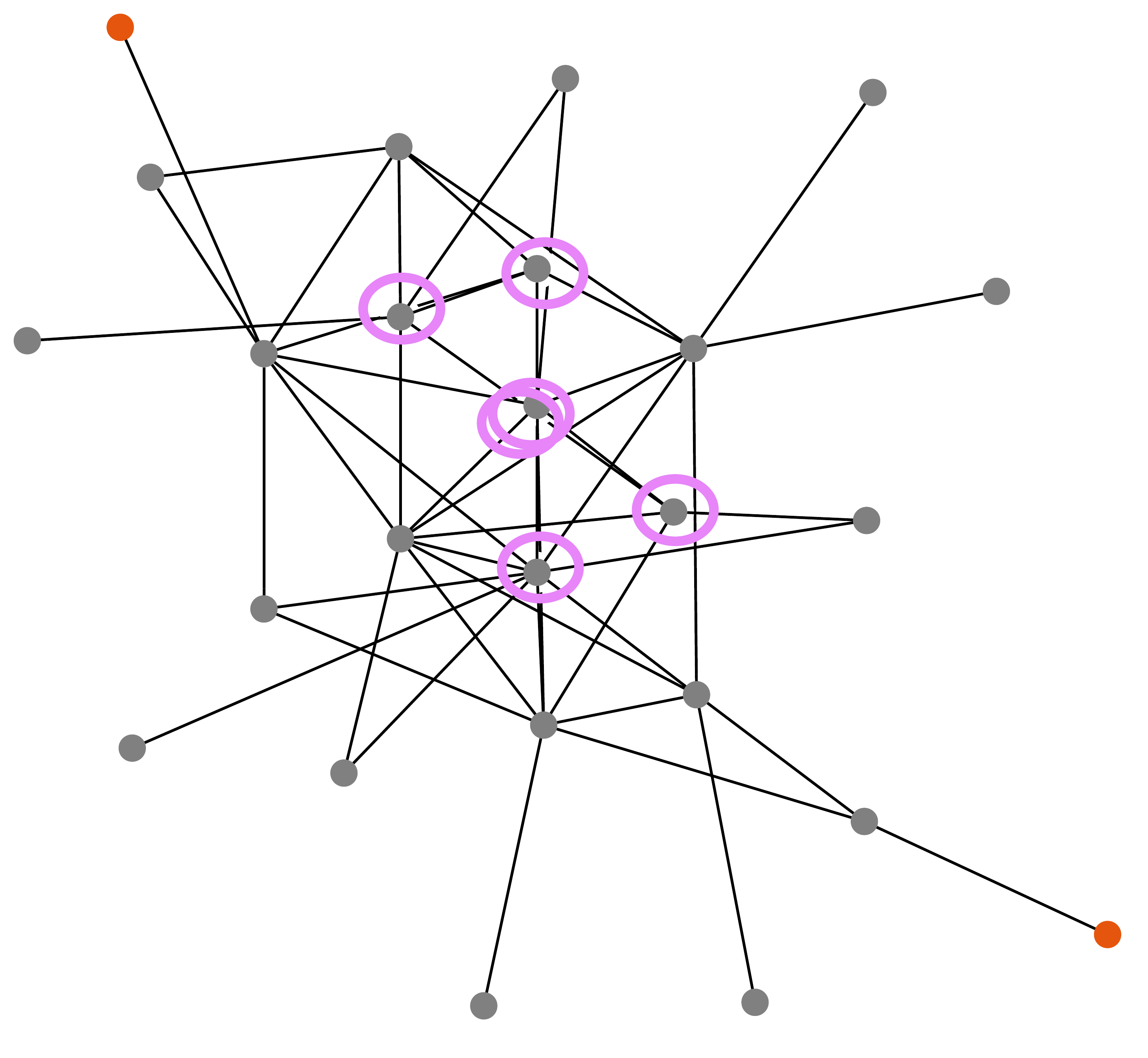}\label{fig:gLN}}
    \subfigure[Link-link crossings on the shortest paths]{\includegraphics[width=0.3\linewidth]{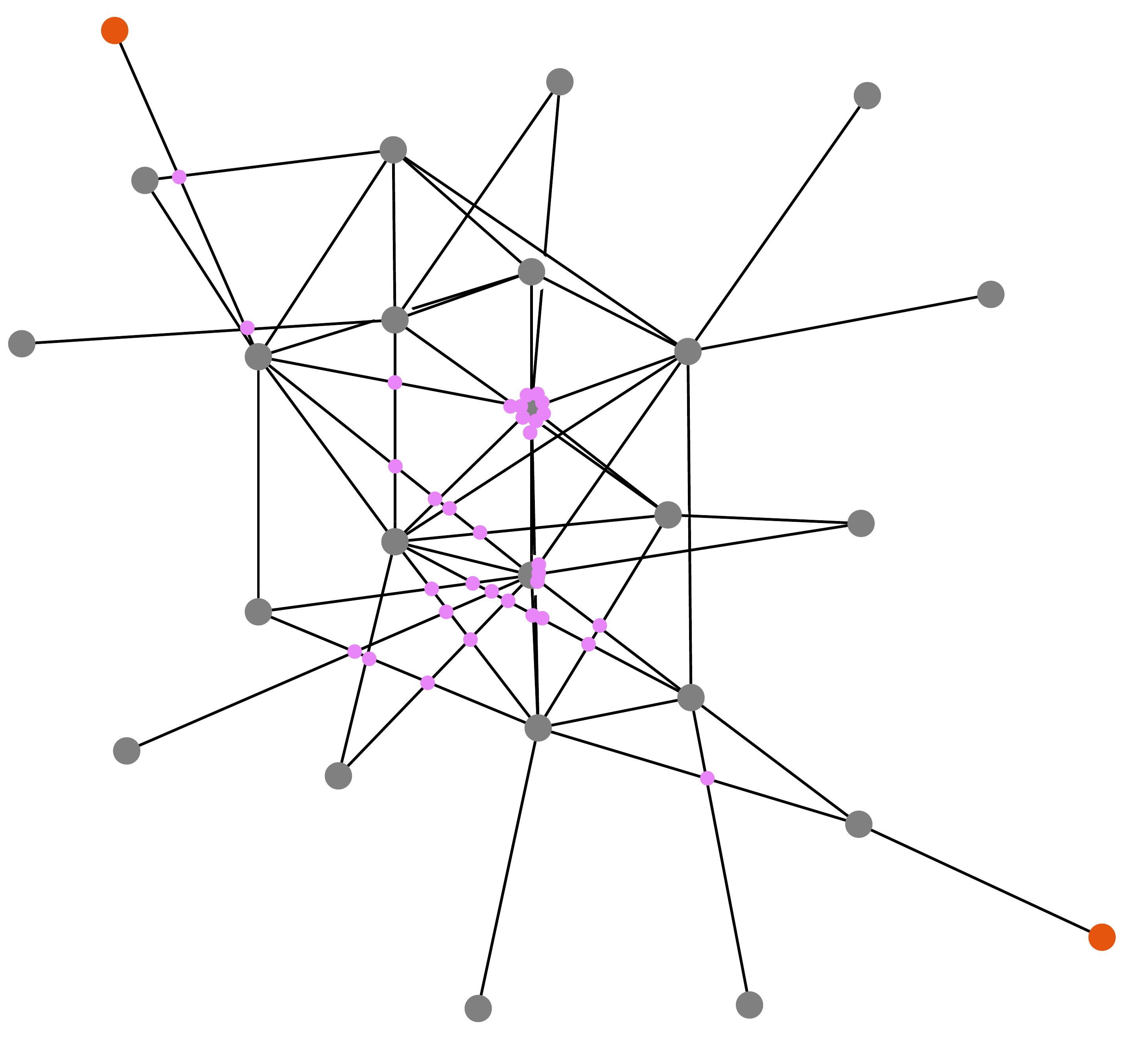}\label{fig:sLL}}
    \subfigure[Node-link crossings on the shortest paths]{\includegraphics[width=0.3\linewidth]{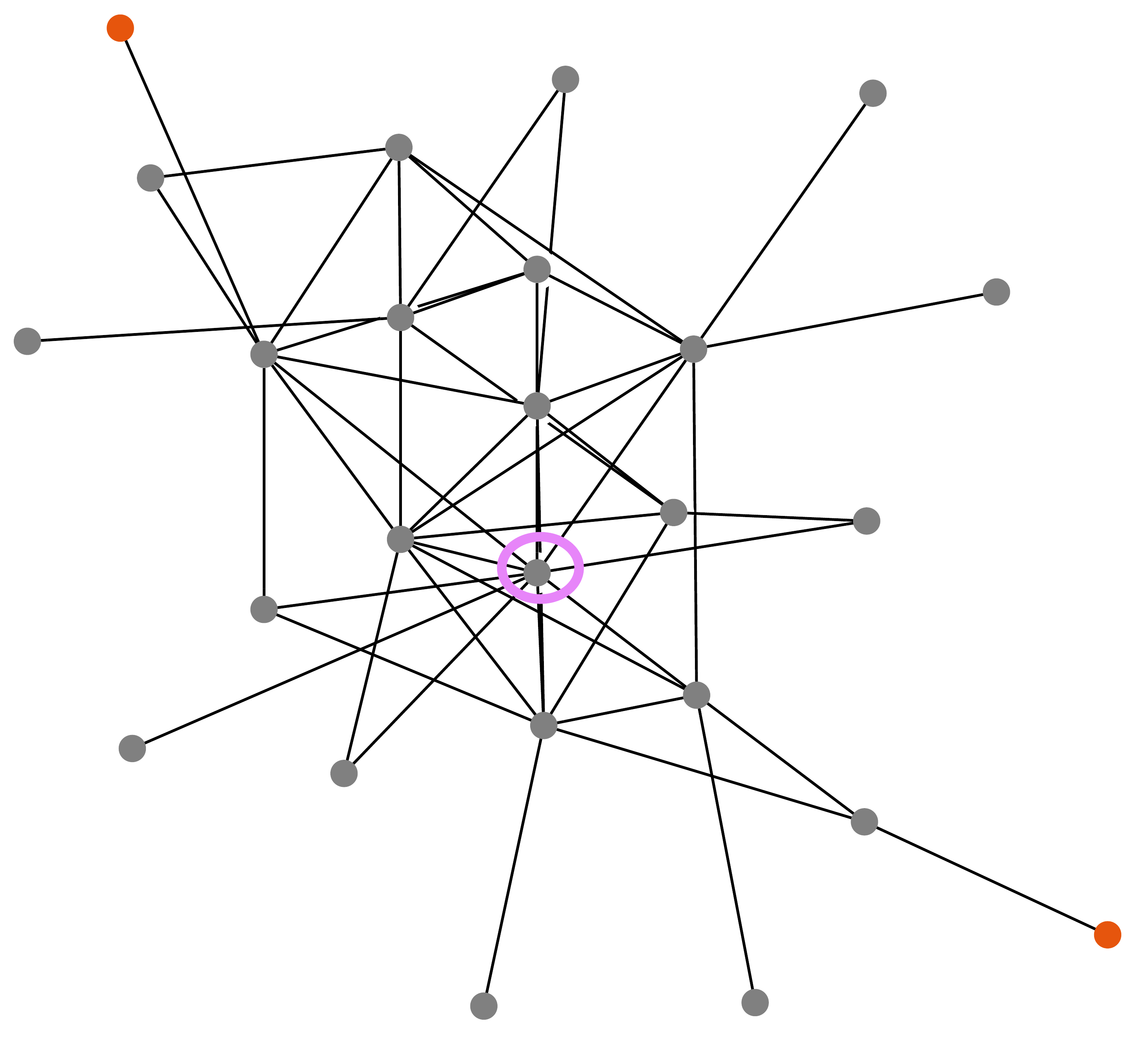}\label{fig:sLN}}
    
    \subfigure[Euclidean distance of the shortest paths]{\includegraphics[width=0.3\linewidth]{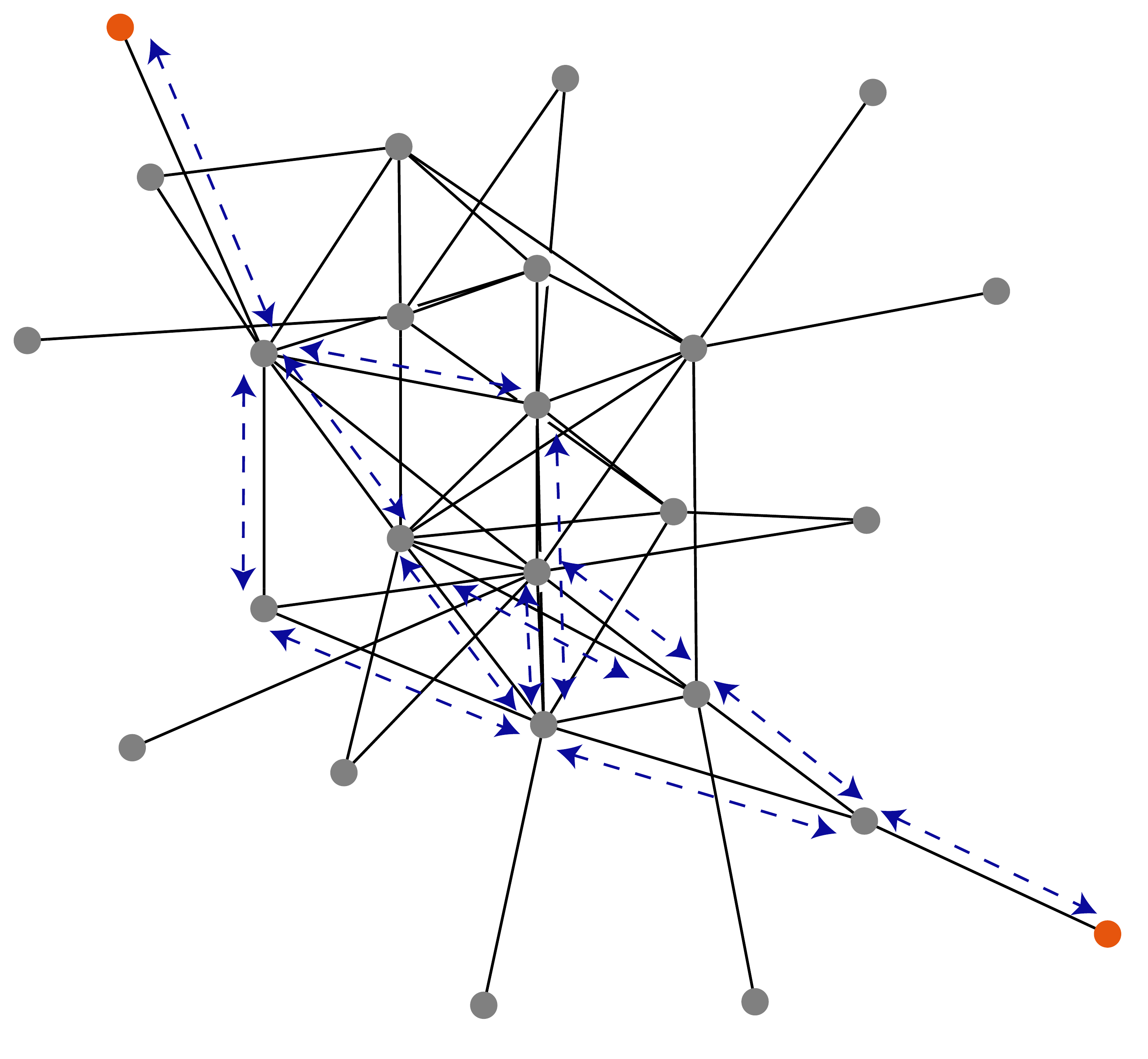}\label{fig:sEuc}}
    \subfigure[Geodesic path deviation]{\includegraphics[width=0.3\linewidth]{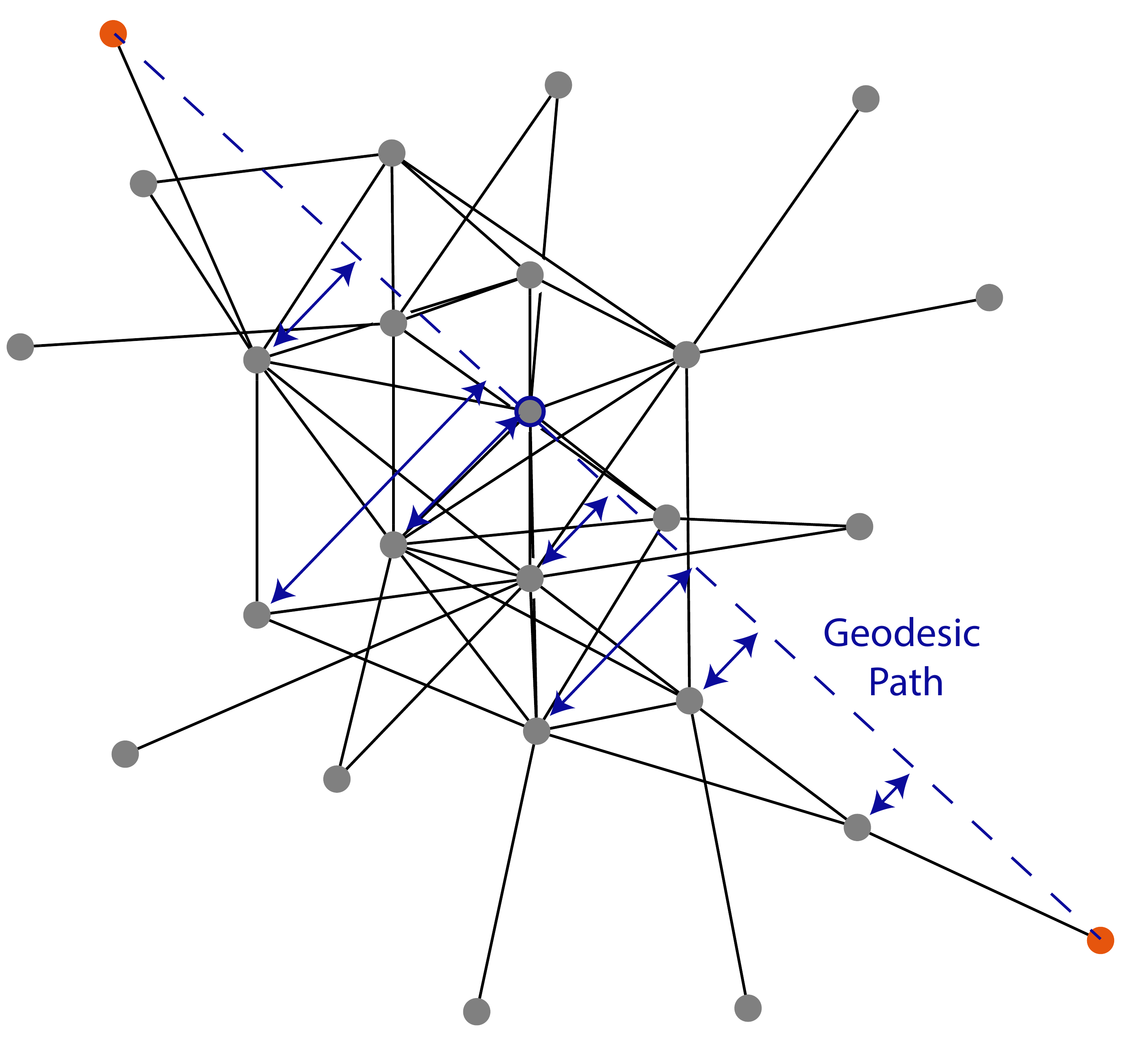}\label{fig:sEq}}
    \subfigure[Degrees of nodes on the shortest paths]{\includegraphics[width=0.3\linewidth]{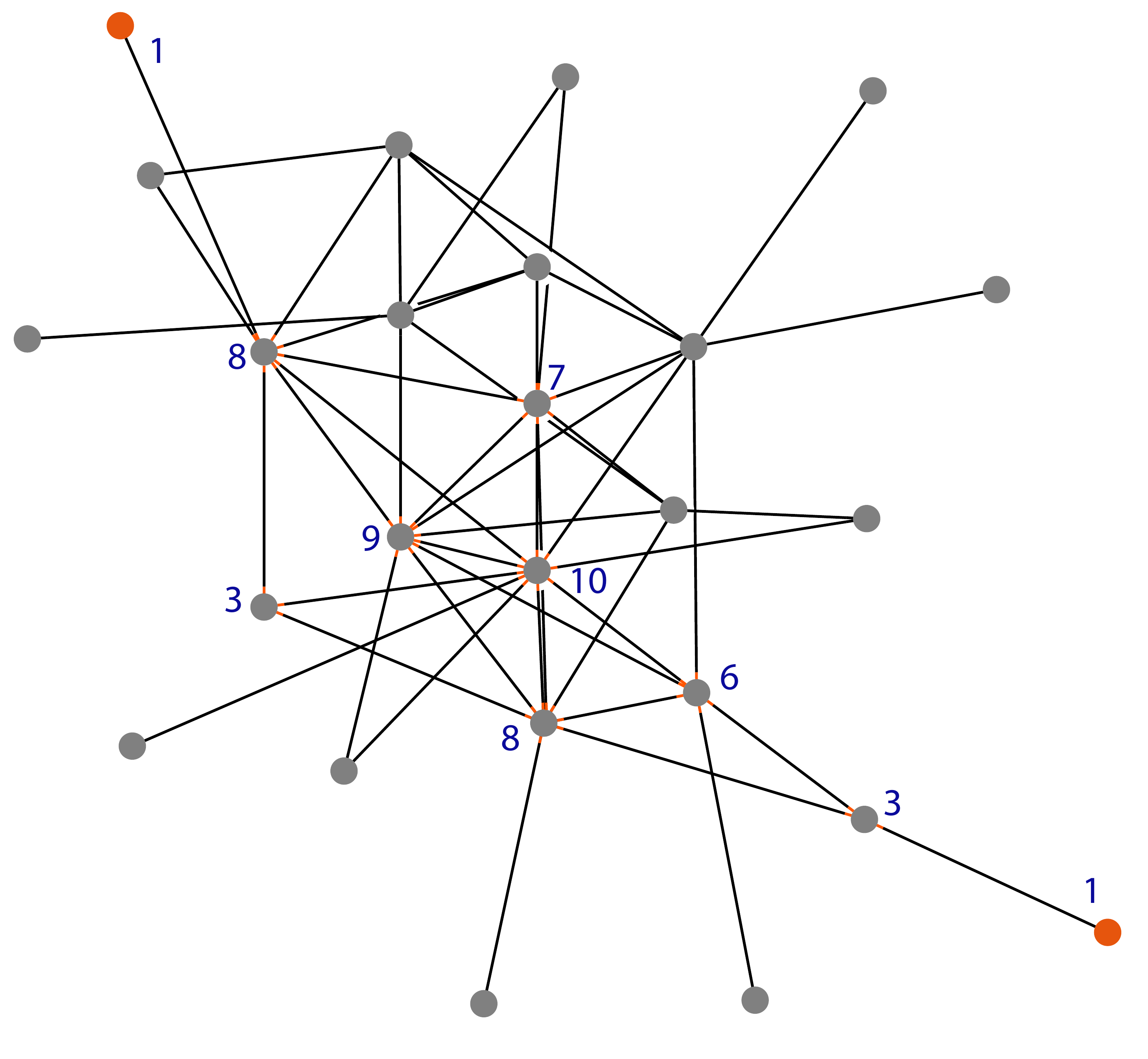}\label{fig:sDeg}}
    \subfigure[Crossing angles on the shortest paths]{\includegraphics[width=0.3\linewidth]{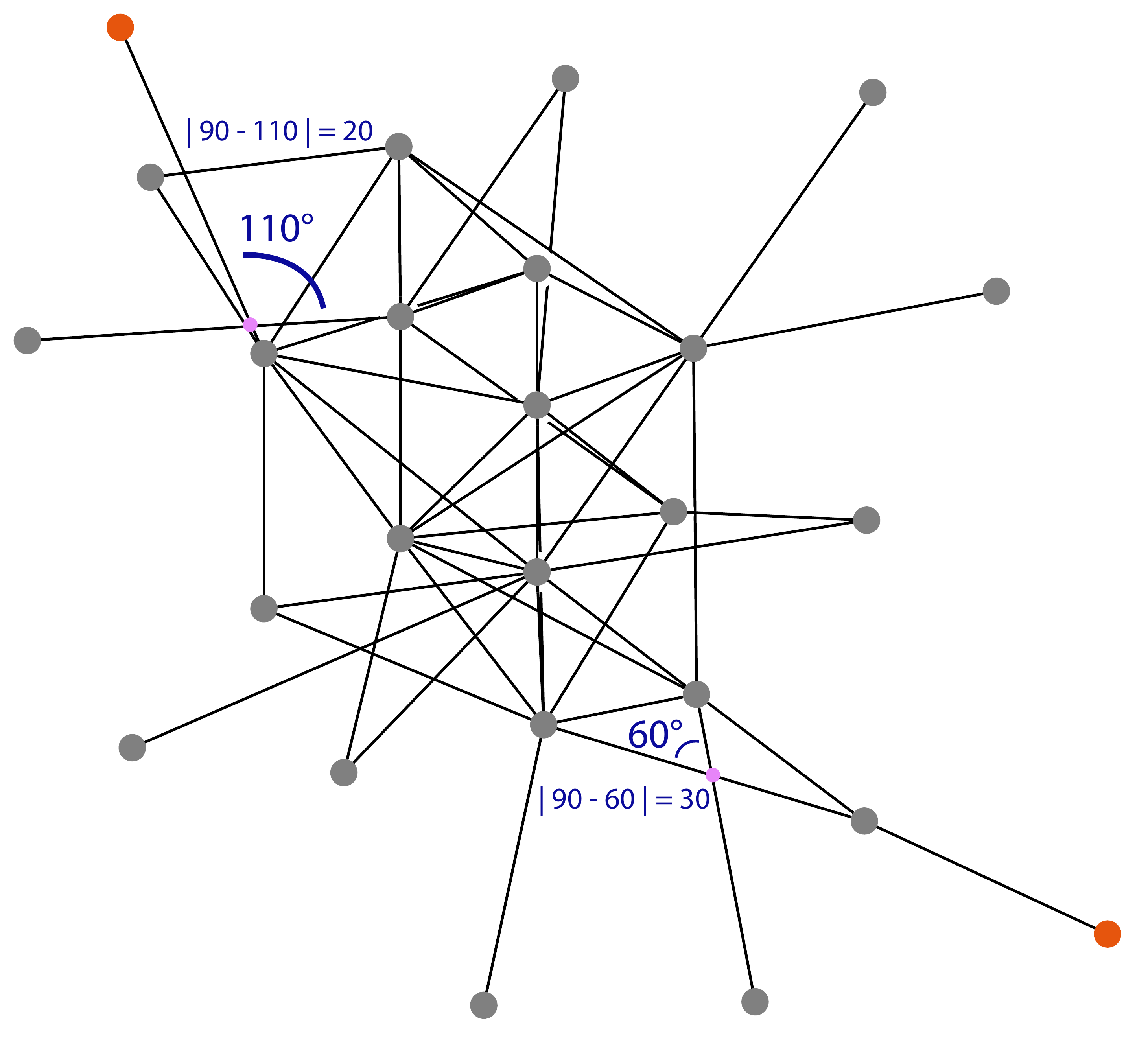}\label{fig:sCA}}
    \subfigure[Turning angles on the shortest paths]{\includegraphics[width=0.3\linewidth]{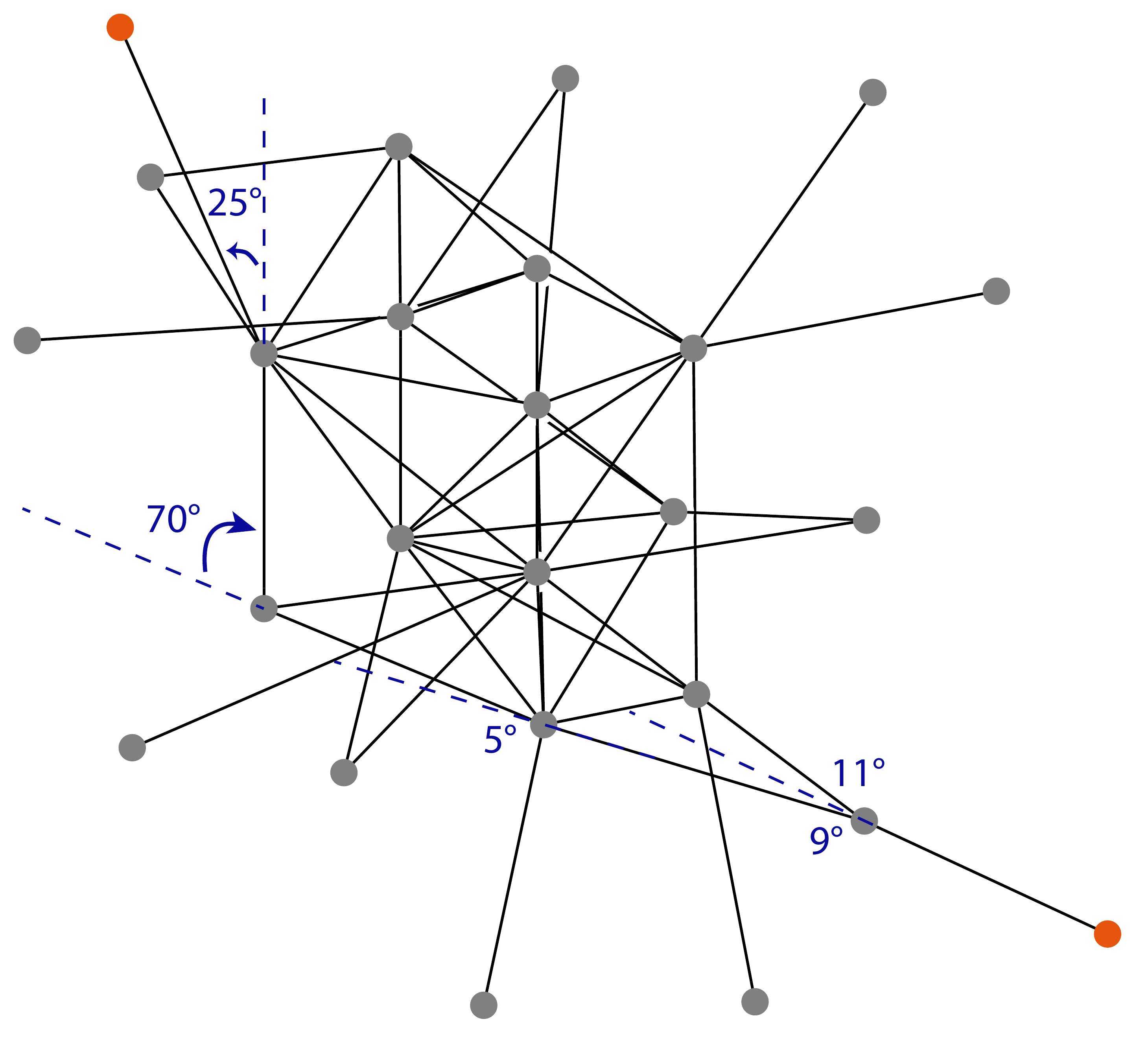}\label{fig:sTA}}
    
    \caption{Features for a sample network used in the study with 25 nodes and 50 edges (density 2). \label{fig:metrics}}
    \end{figure*}

Figure~\ref{fig:metrics} shows examples where some of these features are highlighted. The graph shown in the examples is from our study corpus. It has 6 possible shortest paths (Fig.~\ref{fig:sp}), each with 4 intermediate nodes. The nodes on the shortest paths have an accumulated degree of 56 (Fig.~\ref{fig:sDeg}, average = 28.33/path). The node-link diagram representing the graph has 57 link-link crossings (Fig.~\ref{fig:gLL}) and 6 node-link crossings (Fig.~\ref{fig:gLN}). There are 33 link-link crossings (Fig.~\ref{fig:sLL}, average = 8.33/path) and 1 node-link crossing (Fig.~\ref{fig:sLN}, average = 0.17/path) on the shortest paths. The sum of the Euclidean distance of the links on the shortest paths is 2803.66 (Fig.~\ref{fig:sEuc}, average = 1067.47/path). The sum of the distance between the nodes on the shortest paths from the Geodesic path is 874.46 (Fig.~\ref{fig:sEq}, average = 415.66/path). The sum of the penalty for small angles of link-link crossings on the shortest paths is 925.92 (Fig.~\ref{fig:sCA}, average = 217.08/path). Lastly, the sum of the turning angles on the shortest paths is 426.72 degrees (Fig.~\ref{fig:sTA}, average = 105.58 degrees).

\begin{figure}[t!]
\centering
\includegraphics[width=\columnwidth]{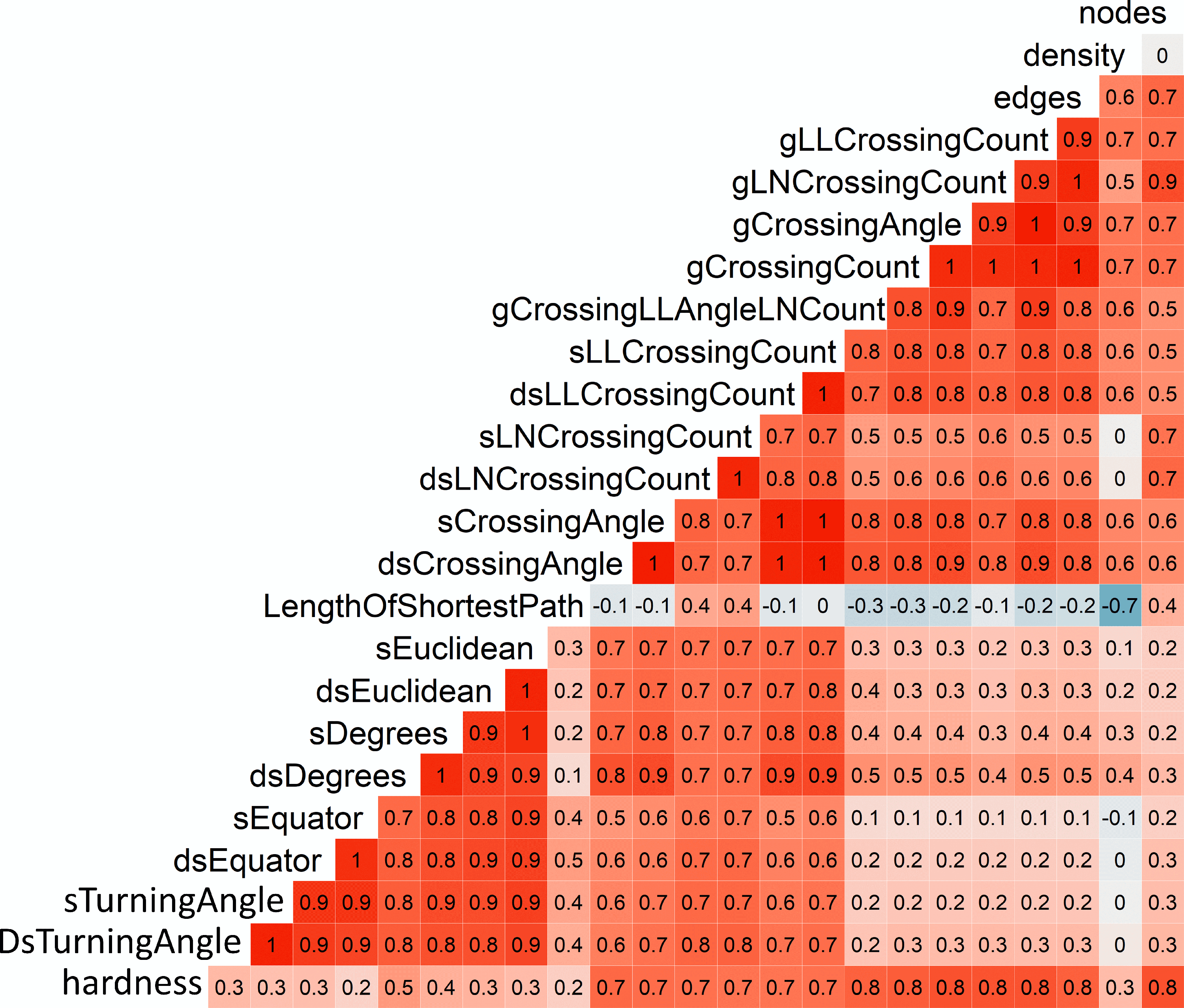}
\caption{Correlation matrix of graph metrics.\label{fig:graph-metrics-correlation}}
\end{figure}

\subsection{Analysis}
As a first step we computed the correlation between these different measures and also with task hardness. See Figure~\ref{fig:graph-metrics-correlation}. As one would expect the different measures of the same basic feature are often closely correlated. Thus, for instance the measures of crossings and crossing angle are highly correlated. Furthermore, as the graph grows the number of node-link and link-link crossings increase. We also see that measures of the Euclidean length of the shortest path(s), its straightness and the degree of the nodes on it are all highly correlated.

We then built multi-level linear models to understand how the above graph features influence task hardness (a similar approach was employed in~\cite{ware2002cognitive}).  This exploratory study used \emph{all-subsets} methods to consider different combinations of features as predictors. 
Following Field \emph{et al.}~\cite[Chap~7.9]{field2012discovering}, we only considered models meeting the following assumptions:
\begin{itemize}
	\item Limited multicollinearity; we used the VIF statistics to calculate VIF values and we considered $VIF<5$ as meet the requirement.
	\item Independence; we used the Durbin-Watson test and considered   models within the range of $[1.5, 2.5]$ as meeting the requirement.
	\item Homoscedasticity (means that residuals at each level of the predictors should have the same variance): different transformations were applied to different graph metrics before building the linear models to meet this requirement:
	
	\noindent\textbf{log transformation}: gLLCrossingCount, gCrossingAngle, sLLCrossingCount, sCrossingAngle, dsCrossingAngle, sEuclidean, dsEuclidean, sDegrees and dsDegrees.
	
	\noindent\textbf{square root transformation}: gLNCrossingCount, dsLNCrossingCount, sEquator, dsEquator, sTurningAngle, dsTurningAngle.
	
	\noindent\textbf{4th square root transformation:} gCrossingCount, sLNCrossingCount and dsLLCrossingCount.
	
	Density was excluded from modelling as the transformed values still did not meet the homoscedasticity assumption.
\end{itemize}
We also normalised each input factor to 0--1 range before modelling.

We report representative models for different number of predictors.  Those for one predictor are, of course, the features most highly correlated with task hardness. The top 12 models are given in Table~\ref{tab:linear-model-1}. What we see is that the number of nodes and global measures of crossing angle or crossing count are the best predictors, followed by number of edges and measures of crossing angle and count for the shortest path. Other features associated with the shortest path such as its length or straightness are poor predictors of task hardness.

The top 12 models for two predictors are given in Table~\ref{tab:linear-model-2}. Here we see that the best model combines global measures of crossing angle or crossing count with the length of the shortest path. The next best combine global measures of crossing angle or crossing count with measures of the number of nodes or graph density.
These are followed by predictors combining global measures of crossings with local measures of crossings on the shortest paths. 

We also considered models with three predictors but none had sufficient extra explanatory power to warrant the use of a third predictor.

\begin{table}
	\centering
	\includegraphics[width=\columnwidth]{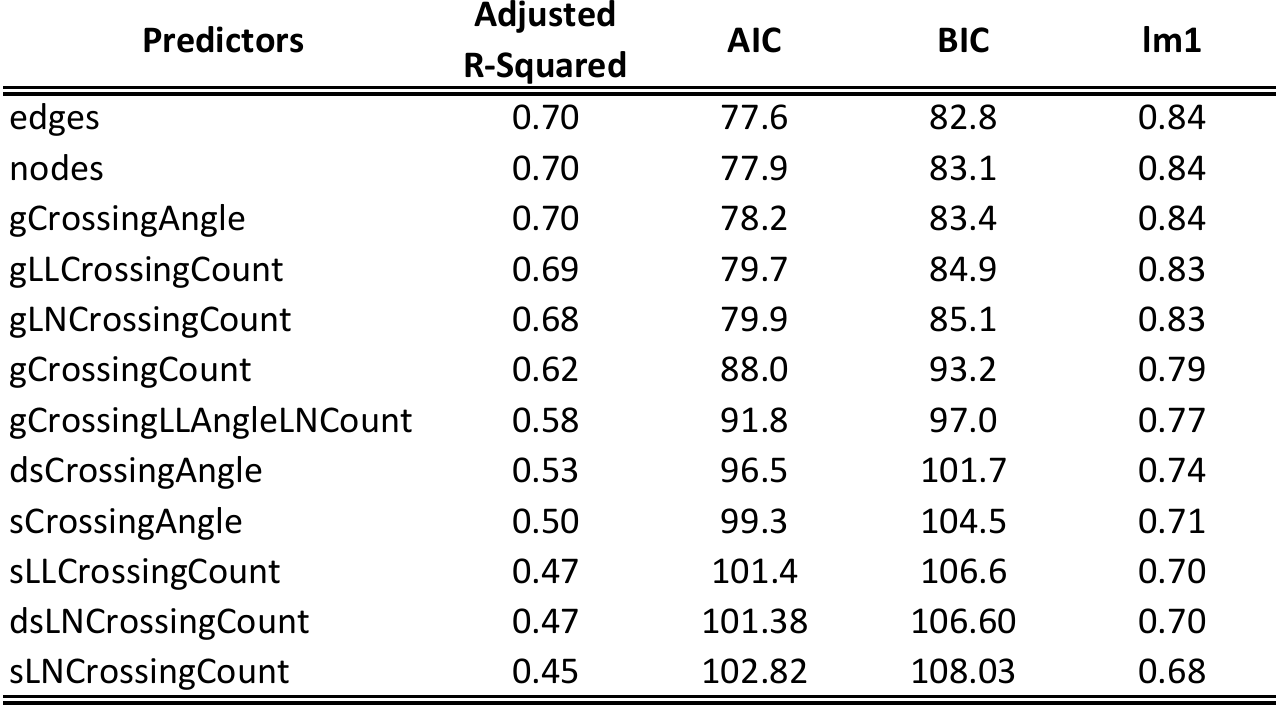}
	\caption{Top 12 linear models with one predictor.}
	\label{tab:linear-model-1}
\end{table}

\begin{table}
	\centering
	\includegraphics[width=\columnwidth]{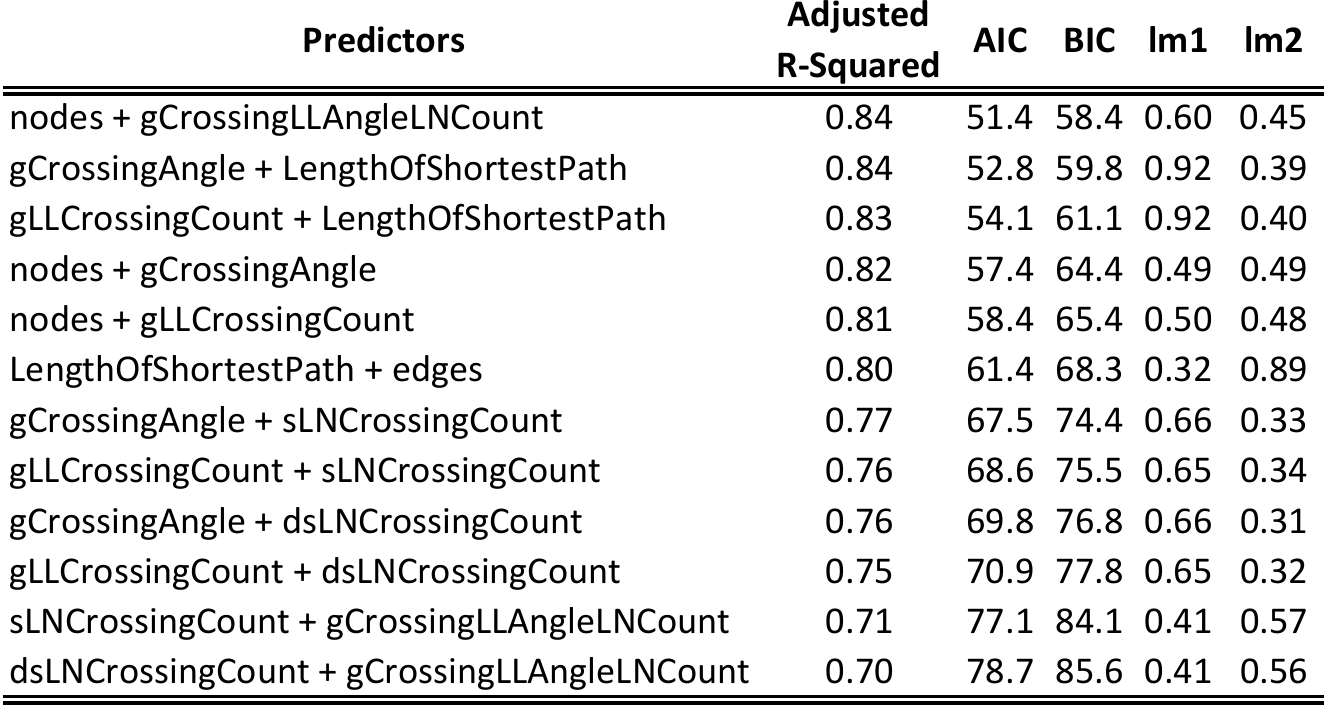}
	\caption{Top 12 linear models with two predictors.}
	\label{tab:linear-model-2}
\end{table}

What we see is that the global graph features are much better predictors of task hardness than features of the shortest path(s). This is a little surprising, likely reflecting that the participants looked at many more paths than the shortest one or because in the stimuli the shortest path was deliberately constructed to run from one side of the graph to the other. We also see that the best combination of two predictors utilise global measures of crossings with either the number of nodes or the number of nodes on the shortest path.

The most similar study is that of~\cite{ware2002cognitive}. They also studied the effect of different graph features on the difficulty of finding the shortest path. They evaluated the effect of shortest path length (number of nodes) and its Euclidean length, shortest path straightness, degree of nodes on shortest path, number of crossings and average crossing angles on shortest path, as well as the global number of crossings.  They did not vary the number of nodes in the stimuli or consider density or number of edges. They found that the two best predictors were shortest path length and straightness of the path. In particular they found that the global number of crossings was not a good predictor but that the number of crossings on the shortest path was. This contrasts with our finding that global predictors such as number of nodes or number of crossings are more influential than number of crossings or straightness of the shortest path. We believe this is because the graphs used in~\cite{ware2002cognitive} were small---only 42 nodes---and relatively sparse with only a few crossings. Consequently the task was much easier: 93\% of responses were correct. We suspect that this means that the participants quickly found the shortest path and so its features dominated, while in our harder experiment participants considered many other paths to the shortest path and so global features were more important.

\section{Limitations and Future Work}
\label{sec:limit}

Our study had a number of limitations. The first is that it was restricted to scale-free graphs and that we used a particular layout algorithm. While we believe that our results apply generally to node-link diagrams, further studies are required to validate this. 

W\deleted{Another limitation was that w}e considered only one task: finding the shortest path between two nodes. We believe that this complex task is representative of a wide-range of path following tasks and that it involves a variety `sub-tasks', such as dis-ambiguating edges, inspecting neighbours, remembering previously inspected nodes, browsing through paths, and so on. Nonetheless other tasks should be considered in future work.

We also realised a number of limitations of the study with respect to our measurement and analysis of the physiological data. Measurements of pupil dilation are sensitive to illumination~\cite{beatty2000pupillary}. A limitation of this study was not  considering the effect of the stimuli on illumination. With larger or denser graphs the screen is slightly darker, reducing illumination and so  increasing pupil dilation. This could potentially explain some of the increase in pupil dilation as task difficulty increased. However, we believe the impact was minimal as the study was conducted in \replaced{a}{an internal} well-lit office and we actually see that pupil dilation decreased when the stimuli became sufficiently difficult. 

Whilst the brain activity patterns revealed in the EEG data accord with the limited available literature, these results should be viewed cautiously. Not only  was there a significant level of noise in the data but the EEG results are likely to be more nuanced. In particular, we ignored individual strategy differences or spatial abilities which are likely to significantly impact on the brain regions used in the task. \replaced{While participants were instructed not to use the mouse while completing the task, a few participants began to use the mouse to trace over the path before being asked not to. This may have resulted in a greater amount of motor, and more importantly, pre-motor cortex activity on the opposite brain hemisphere to the hand being used. This could have resulted in minor differences in brain activity between the easy and hard stimuli in left frontal-central regions, if the right hand was used more with the mouse. However we see little indication of this. Future studies, should take steps  to ensure that participants are not able to use the mouse.}{For example, in this study, some participants used the mouse to trace over the path as this could help with the step-by-step comparison strategy mentioned earlier in the discussion. This may have resulted in a greater amount of motor, and more importantly, pre-motor cortex activity on the opposite brain hemisphere to the hand being used. This could have resulted in differences in brain activity between the easy and hard stimuli in left frontal-central regions, if the right hand was used more with the mouse, and even more so if only on the harder stimuli. In future studies, steps should be taken to ensure that participants are not able to use the mouse.}

It is also important to note the limitations of EEG analysis. While it gives a broad indication of brain activity it is not possible to confidently point out detailed brain regions from our results; source localisation techniques~\cite{pascual-marqui_standardized_2002} are required to allow certainty.

\section{Conclusion}
\label{sec:conc}

We have explored the perceptual limitations of node-link diagrams for a representative connectivity task, finding the shortest path between two nodes. We found that the usefulness of node-link diagrams  rapidly deteriorates as the number of nodes and edges increases. For 
small-world graphs with 50 or more nodes and a density (ratio of edges to nodes) of 6, participants were unable to correctly answer in more than half of the trials.  This was also the case for graphs with a density of 2 and more than 100 nodes.

To the best of our knowledge this is the first study to consider physiological measures of cognitive load (EEG, pupil dilation and heart rate variation) for a network visualisation task. We found that these measures of load initially increase with task hardness but then decrease, presumably because participants give up. The analysis of EEG data was particularly revealing, indicating that  the left frontal, right centro-parietal and left parieto-occipital regions display increased cognitive load for our task. Trace activation was also found in the right frontal region.
We hope that our experience  will inform future visualisation researchers who also wish to use  physiological measures to reveal cognitive load for other kinds of visualisation tasks. 

We also explored the effects of global network layout features such as size or number of crossings and features of the shortest path such as length or straightness on task difficulty. We found that the global measures such as number of crossings had a greater impact than features of the shortest path such as straightness. This is in contrast to an earlier study of Ware~\cite{ware2002cognitive} and may reflect the harder stimuli used in our study.

Our results can guide visualisation designers when creating visualisations that must scale to larger graph data (e.g.,\ setting limits on neighbourhood size in overview-and-detail techniques using node-link diagrams for detail). We also hope this work stimulates development of new techniques that demonstrably scale to larger, more complex networks such as summary representations~\cite{yoghourdjian2018graph}.

\acknowledgments{
The authors wish to acknowledge the support of the Australian Research Council (ARC) through DP140100077. 
Yalong Yang was partially supported by a Harvard Physical Sciences and Engineering Accelerator Award. 
We also wish to thank all our participants for their time and our reviewers for their comments and feedback.
}

\bibliographystyle{abbrv-doi-hyperref}

\bibliography{workload}

\begin{thebibliography}{10}

\bibitem{gtec}
{g.tec: http://gtec.at}.

\bibitem{hrvlogger}
{Heart Rate Variability Logger:
  https://www.marcoaltini.com/blog/heart-rate-variability-logger-app-details}.

\bibitem{tobii}
{Tobii Pro: http://tobiipro.com}.

\bibitem{webcola}
{webcola: https://ialab.it.monash.edu/webcola/}.

\bibitem{albert2005scale}
R.~Albert.
\newblock Scale-free networks in cell biology.
\newblock {\em Journal of cell science}, 118(21):4947--4957, 2005.

\bibitem{anderson2011user}
E.~W. Anderson, K.~C. Potter, L.~E. Matzen, J.~F. Shepherd, G.~A. Preston, and
  C.~T. Silva.
\newblock A user study of visualization effectiveness using eeg and cognitive
  load.
\newblock In {\em Computer Graphics Forum}, vol.~30, pp. 791--800. Wiley Online
  Library, 2011.

\bibitem{antonenko2010using}
P.~Antonenko, F.~Paas, R.~Grabner, and T.~Van~Gog.
\newblock Using electroencephalography to measure cognitive load.
\newblock {\em Educational Psychology Review}, 22(4):425--438, 2010.

\bibitem{archambault2010readability}
D.~Archambault, H.~C. Purchase, and B.~Pinaud.
\newblock The readability of path-preserving clusterings of graphs.
\newblock In {\em Computer Graphics Forum}, vol.~29, pp. 1173--1182. Wiley
  Online Library, 2010.

\bibitem{barabasi1999emergence}
A.-L. Barab{\'a}si and R.~Albert.
\newblock Emergence of scaling in random networks.
\newblock {\em science}, 286(5439):509--512, 1999.

\bibitem{beatty2000pupillary}
J.~Beatty, B.~Lucero-Wagoner, et~al.
\newblock The pupillary system.
\newblock {\em Handbook of psychophysiology}, 2:142--162, 2000.

\bibitem{blumenfeld_putting_2011}
\href{https://doi.org/10.1162/jocn.2010.21459}{R.~S. Blumenfeld, C.~M. Parks,
  A.~P. Yonelinas, and C.~Ranganath}.
\newblock \href{https://doi.org/10.1162/jocn.2010.21459}{Putting the {Pieces}
  {Together}: {The} {Role} of {Dorsolateral} {Prefrontal} {Cortex} in
  {Relational} {Memory} {Encoding}}.
\newblock \href{https://doi.org/10.1162/jocn.2010.21459}{{\em Journal of
  Cognitive Neuroscience}},
  \href{https://doi.org/10.1162/jocn.2010.21459}{23(1):257--265},
  \href{https://doi.org/10.1162/jocn.2010.21459}{Jan. 2011}.
  \href{https://doi.org/10.1162/jocn.2010.21459}
{doi: {{%
10\hspace{.1pt}\discretionary{.}{%
}{.}\hspace{.4pt}1162\discretionary{/}{%
}{/}jocn\hspace{.1pt}\discretionary{.}{%
}{.}\hspace{.4pt}2010\hspace{.1pt}\discretionary{.}{%
}{.}\hspace{.4pt}21459}}}


\bibitem{bratfisch1972perceived}
O.~Bratfisch et~al.
\newblock Perceived item-difficulty in three tests of intellectual performance
  capacity.
\newblock 1972.

\bibitem{cabeza_cognitive_2012}
\href{https://doi.org/10.1016/j.tics.2012.04.008}{R.~Cabeza, E.~Ciaramelli, and
  M.~Moscovitch}.
\newblock \href{https://doi.org/10.1016/j.tics.2012.04.008}{Cognitive
  contributions of the ventral parietal cortex: an integrative theoretical
  account}.
\newblock \href{https://doi.org/10.1016/j.tics.2012.04.008}{{\em Trends in
  Cognitive Sciences}},
  \href{https://doi.org/10.1016/j.tics.2012.04.008}{16(6):338--352},
  \href{https://doi.org/10.1016/j.tics.2012.04.008}{June 2012}.
  \href{https://doi.org/10.1016/j.tics.2012.04.008}
{doi: {{%
10\hspace{.1pt}\discretionary{.}{%
}{.}\hspace{.4pt}1016\discretionary{/}{%
}{/}j\hspace{.1pt}\discretionary{.}{%
}{.}\hspace{.4pt}tics\hspace{.1pt}\discretionary{.}{%
}{.}\hspace{.4pt}2012\hspace{.1pt}\discretionary{.}{%
}{.}\hspace{.4pt}04\hspace{.1pt}\discretionary{.}{%
}{.}\hspace{.4pt}008}}}


\bibitem{cabeza_imaging_2000}
\href{https://doi.org/10.1162/08989290051137585}{R.~Cabeza and L.~Nyberg}.
\newblock \href{https://doi.org/10.1162/08989290051137585}{Imaging {Cognition}
  {II}: {An} {Empirical} {Review} of 275 {PET} and {fMRI} {Studies}}.
\newblock \href{https://doi.org/10.1162/08989290051137585}{{\em Journal of
  Cognitive Neuroscience}},
  \href{https://doi.org/10.1162/08989290051137585}{12(1):1--47},
  \href{https://doi.org/10.1162/08989290051137585}{Jan. 2000}.
  \href{https://doi.org/10.1162/08989290051137585}
{doi: {{%
10\hspace{.1pt}\discretionary{.}{%
}{.}\hspace{.4pt}1162\discretionary{/}{%
}{/}08989290051137585}}}


\bibitem{castro2020validating}
L.~J. Castro-Meneses, J.-L. Kruger, and S.~Doherty.
\newblock Validating theta power as an objective measure of cognitive load in
  educational video.
\newblock {\em Educational Technology Research and Development},
  68(1):181--202, 2020.

\bibitem{chandler1991cognitive}
P.~Chandler and J.~Sweller.
\newblock Cognitive load theory and the format of instruction.
\newblock {\em Cognition and instruction}, 8(4):293--332, 1991.

\bibitem{cohen2013statistical}
J.~Cohen.
\newblock {\em Statistical power analysis for the behavioral sciences}.
\newblock Academic press, 2013.

\bibitem{costello_best_2019}
\href{https://doi.org/https://doi.org/10.7275/jyj1-4868}{A.~Costello and
  J.~Osborne}.
\newblock \href{https://doi.org/https://doi.org/10.7275/jyj1-4868}{Best
  practices in exploratory factor analysis: four recommendations for getting
  the most from your analysis}.
\newblock \href{https://doi.org/https://doi.org/10.7275/jyj1-4868}{{\em
  Practical Assessment, Research, and Evaluation}},
  \href{https://doi.org/https://doi.org/10.7275/jyj1-4868}{10(1)},
  \href{https://doi.org/https://doi.org/10.7275/jyj1-4868}{Nov. 2019}.
  \href{https://doi.org/10.7275/jyj1-4868}
{doi: {{%
10\hspace{.1pt}\discretionary{.}{%
}{.}\hspace{.4pt}7275\discretionary{/}{%
}{/}jyj1\discretionary{%
}{-}{-}4868}}}


\bibitem{dan_real_2017}
\href{https://doi.org/10.5281/zenodo.3554719}{A.~Dan and M.~Reiner}.
\newblock \href{https://doi.org/10.5281/zenodo.3554719}{Real {Time} {EEG}
  {Based} {Measurements} of {Cognitive} {Load} {Indicates} {Mental} {States}
  {During} {Learning}}.
\newblock \href{https://doi.org/10.5281/zenodo.3554719}{{\em JEDM {\textbar}
  Journal of Educational Data Mining}},
  \href{https://doi.org/10.5281/zenodo.3554719}{9(2):31--44},
  \href{https://doi.org/10.5281/zenodo.3554719}{Dec. 2017}.
\newblock \href{https://doi.org/10.5281/zenodo.3554719}{Number: 2}.
  \href{https://doi.org/10.5281/zenodo.3554719}
{doi: {{%
10\hspace{.1pt}\discretionary{.}{%
}{.}\hspace{.4pt}5281\discretionary{/}{%
}{/}zenodo\hspace{.1pt}\discretionary{.}{%
}{.}\hspace{.4pt}3554719}}}


\bibitem{dawson2015search}
J.~Q. Dawson, T.~Munzner, and J.~McGrenere.
\newblock A search-set model of path tracing in graphs.
\newblock {\em Information Visualization}, 14(4):308--338, 2015.

\bibitem{de1996measurement}
D.~De~Waard.
\newblock {\em The measurement of drivers' mental workload}.
\newblock Groningen University, Traffic Research Center Netherlands, 1996.

\bibitem{dunne2013motif}
C.~Dunne and B.~Shneiderman.
\newblock Motif simplification: improving network visualization readability
  with fan, connector, and clique glyphs.
\newblock In {\em Proceedings of the SIGCHI Conference on Human Factors in
  Computing Systems}, pp. 3247--3256. ACM, 2013.

\bibitem{eick2002visual}
S.~G. Eick and A.~F. Karr.
\newblock Visual scalability.
\newblock {\em Journal of Computational and Graphical Statistics},
  11(1):22--43, 2002.

\bibitem{faloutsos1999power}
M.~Faloutsos, P.~Faloutsos, and C.~Faloutsos.
\newblock On power-law relationships of the internet topology.
\newblock In {\em ACM SIGCOMM computer communication review}, vol.~29, pp.
  251--262. ACM, 1999.

\bibitem{farkas2002networks}
I.~Farkas, I.~Der{\'e}nyi, H.~Jeong, Z.~Neda, Z.~Oltvai, E.~Ravasz,
  A.~Schubert, A.-L. Barab{\'a}si, and T.~Vicsek.
\newblock Networks in life: Scaling properties and eigenvalue spectra.
\newblock {\em Physica A: Statistical Mechanics and its Applications},
  314(1-4):25--34, 2002.

\bibitem{field2012discovering}
A.~Field, J.~Miles, and Z.~Field.
\newblock {\em Discovering statistics using R}.
\newblock Sage publications, 2012.

\bibitem{gaser_brain_2003}
\href{https://doi.org/10.1523/JNEUROSCI.23-27-09240.2003}{C.~Gaser and
  G.~Schlaug}.
\newblock \href{https://doi.org/10.1523/JNEUROSCI.23-27-09240.2003}{Brain
  {Structures} {Differ} between {Musicians} and {Non}-{Musicians}}.
\newblock \href{https://doi.org/10.1523/JNEUROSCI.23-27-09240.2003}{{\em The
  Journal of Neuroscience}},
  \href{https://doi.org/10.1523/JNEUROSCI.23-27-09240.2003}{23(27):9240--9245},
  \href{https://doi.org/10.1523/JNEUROSCI.23-27-09240.2003}{Oct. 2003}.
  \href{https://doi.org/10.1523/JNEUROSCI.23-27-09240.2003}
{doi: {{%
10\hspace{.1pt}\discretionary{.}{%
}{.}\hspace{.4pt}1523\discretionary{/}{%
}{/}JNEUROSCI\hspace{.1pt}\discretionary{.}{%
}{.}\hspace{.4pt}23\discretionary{%
}{-}{-}27\discretionary{%
}{-}{-}09240\hspace{.1pt}\discretionary{.}{%
}{.}\hspace{.4pt}2003}}}


\bibitem{ghoniem2004comparison}
M.~Ghoniem, J.-D. Fekete, and P.~Castagliola.
\newblock A comparison of the readability of graphs using node-link and
  matrix-based representations.
\newblock In {\em Information Visualization, 2004. INFOVIS 2004. IEEE Symposium
  on}, pp. 17--24. IEEE, 2004.

\bibitem{greffard2011visual}
N.~Greffard, F.~Picarougne, and P.~Kuntz.
\newblock Visual community detection: An evaluation of 2d, 3d perspective and
  3d stereoscopic displays.
\newblock In {\em International Symposium on Graph Drawing}, pp. 215--225.
  Springer, 2011.

\bibitem{gNautilusInstructions}
{g.tec medical engineering GmbH}.
\newblock {g.Nautilus} wireless biosignal acquisition: Instructions for use,
  Oct 2017.
\newblock V1.16.06.

\bibitem{haapalainen_psycho-physiological_2010}
\href{https://doi.org/10.1145/1864349.1864395}{E.~Haapalainen, S.~Kim, J.~F.
  Forlizzi, and A.~K. Dey}.
\newblock \href{https://doi.org/10.1145/1864349.1864395}{Psycho-physiological
  measures for assessing cognitive load}.
\newblock \href{https://doi.org/10.1145/1864349.1864395}{In {\em Proceedings of
  the 12th {ACM} international conference on {Ubiquitous} computing}},
  \href{https://doi.org/10.1145/1864349.1864395}{pp. 301--310}.
  \href{https://doi.org/10.1145/1864349.1864395}{ACM},
  \href{https://doi.org/10.1145/1864349.1864395}{Copenhagen Denmark},
  \href{https://doi.org/10.1145/1864349.1864395}{Sept. 2010}.
  \href{https://doi.org/10.1145/1864349.1864395}
{doi: {{%
10\hspace{.1pt}\discretionary{.}{%
}{.}\hspace{.4pt}1145\discretionary{/}{%
}{/}1864349\hspace{.1pt}\discretionary{.}{%
}{.}\hspace{.4pt}1864395}}}


\bibitem{huang2007using}
W.~Huang.
\newblock Using eye tracking to investigate graph layout effects.
\newblock In {\em Visualization, 2007. APVIS'07. 2007 6th International
  Asia-Pacific Symposium on}, pp. 97--100. IEEE, 2007.

\bibitem{huang2009measuring}
W.~Huang, P.~Eades, and S.-H. Hong.
\newblock Measuring effectiveness of graph visualizations: A cognitive load
  perspective.
\newblock {\em Information Visualization}, 8(3):139--152, 2009.

\bibitem{huang2008effects}
W.~Huang, S.-H. Hong, and P.~Eades.
\newblock Effects of crossing angles.
\newblock In {\em Visualization Symposium, 2008. PacificVIS'08. IEEE Pacific},
  pp. 41--46. IEEE, 2008.

\bibitem{jacobs_eeg_2006}
\href{https://doi.org/10.1016/j.neuroimage.2006.02.018}{J.~Jacobs, G.~Hwang,
  T.~Curran, and M.~J. Kahana}.
\newblock \href{https://doi.org/10.1016/j.neuroimage.2006.02.018}{{EEG}
  oscillations and recognition memory: {Theta} correlates of memory retrieval
  and decision making}.
\newblock \href{https://doi.org/10.1016/j.neuroimage.2006.02.018}{{\em
  NeuroImage}},
  \href{https://doi.org/10.1016/j.neuroimage.2006.02.018}{32(2):978--987},
  \href{https://doi.org/10.1016/j.neuroimage.2006.02.018}{Aug. 2006}.
  \href{https://doi.org/10.1016/j.neuroimage.2006.02.018}
{doi: {{%
10\hspace{.1pt}\discretionary{.}{%
}{.}\hspace{.4pt}1016\discretionary{/}{%
}{/}j\hspace{.1pt}\discretionary{.}{%
}{.}\hspace{.4pt}neuroimage\hspace{.1pt}\discretionary{.}{%
}{.}\hspace{.4pt}2006\hspace{.1pt}\discretionary{.}{%
}{.}\hspace{.4pt}02\hspace{.1pt}\discretionary{.}{%
}{.}\hspace{.4pt}018}}}


\bibitem{jankun2014scalability}
T.~Jankun-Kelly, T.~Dwyer, D.~Holten, C.~Hurter, M.~N{\"o}llenburg, C.~Weaver,
  and K.~Xu.
\newblock Scalability considerations for multivariate graph visualization.
\newblock In {\em Multivariate Network Visualization}, pp. 207--235. Springer,
  2014.

\bibitem{bakdash2017}
\href{https://doi.org/10.3389/fpsyg.2017.00456}{L.~R.~M. Jonathan Z.~Bakdash}.
\newblock \href{https://doi.org/10.3389/fpsyg.2017.00456}{{Repeated Measures
  Correlation}}.
\newblock \href{https://doi.org/10.3389/fpsyg.2017.00456}{{\em Frontiers in
  Psychology}}, \href{https://doi.org/10.3389/fpsyg.2017.00456}{8:456},
  \href{https://doi.org/10.3389/fpsyg.2017.00456}{2017}.
  \href{https://doi.org/10.3389/fpsyg.2017.00456}
{doi: {{%
10\hspace{.1pt}\discretionary{.}{%
}{.}\hspace{.4pt}3389\discretionary{/}{%
}{/}fpsyg\hspace{.1pt}\discretionary{.}{%
}{.}\hspace{.4pt}2017\hspace{.1pt}\discretionary{.}{%
}{.}\hspace{.4pt}00456}}}


\bibitem{kahn_factor_2006}
\href{https://doi.org/10.1177/0011000006286347}{J.~H. Kahn}.
\newblock \href{https://doi.org/10.1177/0011000006286347}{Factor {Analysis} in
  {Counseling} {Psychology} {Research}, {Training}, and {Practice}:
  {Principles}, {Advances}, and {Applications}}.
\newblock \href{https://doi.org/10.1177/0011000006286347}{{\em The Counseling
  Psychologist}},
  \href{https://doi.org/10.1177/0011000006286347}{34(5):684--718},
  \href{https://doi.org/10.1177/0011000006286347}{Sept. 2006}.
  \href{https://doi.org/10.1177/0011000006286347}
{doi: {{%
10\hspace{.1pt}\discretionary{.}{%
}{.}\hspace{.4pt}1177\discretionary{/}{%
}{/}0011000006286347}}}


\bibitem{kaplan2017neural}
R.~Kaplan, J.~King, R.~Koster, W.~D. Penny, N.~Burgess, and K.~J. Friston.
\newblock The neural representation of prospective choice during spatial
  planning and decisions.
\newblock {\em PLoS biology}, 15(1), 2017.

\bibitem{kaplan_neural_2017}
\href{https://doi.org/10.1371/journal.pbio.1002588}{R.~Kaplan, J.~King,
  R.~Koster, W.~D. Penny, N.~Burgess, and K.~J. Friston}.
\newblock \href{https://doi.org/10.1371/journal.pbio.1002588}{The {Neural}
  {Representation} of {Prospective} {Choice} during {Spatial} {Planning} and
  {Decisions}}.
\newblock \href{https://doi.org/10.1371/journal.pbio.1002588}{{\em PLOS
  Biology}},
  \href{https://doi.org/10.1371/journal.pbio.1002588}{15(1):e1002588},
  \href{https://doi.org/10.1371/journal.pbio.1002588}{Jan. 2017}.
  \href{https://doi.org/10.1371/journal.pbio.1002588}
{doi: {{%
10\hspace{.1pt}\discretionary{.}{%
}{.}\hspace{.4pt}1371\discretionary{/}{%
}{/}journal\hspace{.1pt}\discretionary{.}{%
}{.}\hspace{.4pt}pbio\hspace{.1pt}\discretionary{.}{%
}{.}\hspace{.4pt}1002588}}}


\bibitem{keller2006matrices}
R.~Keller, C.~M. Eckert, and P.~J. Clarkson.
\newblock Matrices or node-link diagrams: which visual representation is better
  for visualising connectivity models?
\newblock {\em Information Visualization}, 5(1):62--76, 2006.

\bibitem{klimesch_eeg_1999}
\href{https://doi.org/10.1016/S0165-0173(98)00056-3}{W.~Klimesch}.
\newblock \href{https://doi.org/10.1016/S0165-0173(98)00056-3}{{EEG} alpha and
  theta oscillations reflect cognitive and memory performance: a review and
  analysis}.
\newblock \href{https://doi.org/10.1016/S0165-0173(98)00056-3}{{\em Brain
  Research Reviews}},
  \href{https://doi.org/10.1016/S0165-0173(98)00056-3}{29(2-3):169--195},
  \href{https://doi.org/10.1016/S0165-0173(98)00056-3}{Apr. 1999}.
  \href{https://doi.org/10.1016/S0165-0173(98)00056-3}
{doi: {{%
10\hspace{.1pt}\discretionary{.}{%
}{.}\hspace{.4pt}1016\discretionary{/}{%
}{/}S0165\discretionary{%
}{-}{-}0173\discretionary{%
}{(}{(}98\discretionary{)}{%
}{)}00056\discretionary{%
}{-}{-}3}}}


\bibitem{kobourov2014crossings}
S.~G. Kobourov, S.~Pupyrev, and B.~Saket.
\newblock Are crossings important for drawing large graphs?
\newblock In {\em International Symposium on Graph Drawing}, pp. 234--245.
  Springer, 2014.

\bibitem{lee2016communities}
A.~Lee and D.~Archambault.
\newblock Communities found by users--not algorithms: Comparing human and
  algorithmically generated communities.
\newblock In {\em Proceedings of the 2016 CHI Conference on Human Factors in
  Computing Systems}, pp. 2396--2400. ACM, 2016.

\bibitem{marner2014gion}
M.~R. Marner, R.~T. Smith, B.~H. Thomas, K.~Klein, P.~Eades, and S.-H. Hong.
\newblock Gion: Interactively untangling large graphs on wall-sized displays.
\newblock In {\em International Symposium on Graph Drawing}, pp. 113--124.
  Springer, 2014.

\bibitem{melancon2006just}
G.~Melancon.
\newblock Just how dense are dense graphs in the real world?: a methodological
  note.
\newblock In {\em Proceedings of the 2006 AVI workshop on BEyond time and
  errors: novel evaluation methods for information visualization}, pp. 1--7.
  ACM, 2006.

\bibitem{moscovich2009topology}
T.~Moscovich, F.~Chevalier, N.~Henry, E.~Pietriga, and J.-D. Fekete.
\newblock Topology-aware navigation in large networks.
\newblock In {\em Proceedings of the SIGCHI Conference on Human Factors in
  Computing Systems}, pp. 2319--2328. ACM, 2009.

\bibitem{nekrasovski2006evaluation}
D.~Nekrasovski, A.~Bodnar, J.~McGrenere, F.~Guimbreti{\`e}re, and T.~Munzner.
\newblock An evaluation of pan \& zoom and rubber sheet navigation with and
  without an overview.
\newblock In {\em Proceedings of the SIGCHI conference on Human Factors in
  computing systems}, pp. 11--20. ACM, 2006.

\bibitem{okoe2018node}
M.~Okoe, R.~Jianu, and S.~G. Kobourov.
\newblock Node-link or adjacency matrices: Old question, new insights.
\newblock {\em IEEE transactions on visualization and computer graphics}, 2018.

\bibitem{oostenveld_fieldtrip_2011}
\href{https://doi.org/10.1155/2011/156869}{R.~Oostenveld, P.~Fries, E.~Maris,
  and J.-M. Schoffelen}.
\newblock \href{https://doi.org/10.1155/2011/156869}{{FieldTrip}: {Open}
  {Source} {Software} for {Advanced} {Analysis} of {MEG}, {EEG}, and {Invasive}
  {Electrophysiological} {Data}}.
\newblock \href{https://doi.org/10.1155/2011/156869}{{\em Computational
  Intelligence and Neuroscience}},
  \href{https://doi.org/10.1155/2011/156869}{2011:1--9},
  \href{https://doi.org/10.1155/2011/156869}{2011}.
  \href{https://doi.org/10.1155/2011/156869}
{doi: {{%
10\hspace{.1pt}\discretionary{.}{%
}{.}\hspace{.4pt}1155\discretionary{/}{%
}{/}2011\discretionary{/}{%
}{/}156869}}}


\bibitem{paas1993efficiency}
F.~G. Paas and J.~J. Van~Merri{\"e}nboer.
\newblock The efficiency of instructional conditions: An approach to combine
  mental effort and performance measures.
\newblock {\em Human factors}, 35(4):737--743, 1993.

\bibitem{pascual-marqui_standardized_2002}
R.~D. Pascual-Marqui.
\newblock Standardized low-resolution brain electromagnetic tomography
  ({sLORETA}): technical details.
\newblock {\em Methods and Findings in Experimental and Clinical Pharmacology},
  24 Suppl D:5--12, 2002.

\bibitem{peck2013using}
E.~M.~M. Peck, B.~F. Yuksel, A.~Ottley, R.~J. Jacob, and R.~Chang.
\newblock Using fnirs brain sensing to evaluate information visualization
  interfaces.
\newblock In {\em Proceedings of the SIGCHI Conference on Human Factors in
  Computing Systems}, pp. 473--482. ACM, 2013.

\bibitem{purchase1997aesthetic}
H.~Purchase.
\newblock Which aesthetic has the greatest effect on human understanding?
\newblock In {\em International Symposium on Graph Drawing}, pp. 248--261.
  Springer, 1997.

\bibitem{purchase1995validating}
H.~C. Purchase, R.~F. Cohen, and M.~James.
\newblock Validating graph drawing aesthetics.
\newblock In {\em International Symposium on Graph Drawing}, pp. 435--446.
  Springer, 1995.

\bibitem{rowe_heart_1998}
\href{https://doi.org/10.1145/274644.274709}{D.~W. Rowe, J.~Sibert, and
  D.~Irwin}.
\newblock \href{https://doi.org/10.1145/274644.274709}{Heart rate variability:
  indicator of user state as an aid to human-computer interaction}.
\newblock \href{https://doi.org/10.1145/274644.274709}{In {\em Proceedings of
  the {SIGCHI} conference on {Human} factors in computing systems - {CHI}
  '98}}, \href{https://doi.org/10.1145/274644.274709}{pp. 480--487}.
  \href{https://doi.org/10.1145/274644.274709}{ACM Press},
  \href{https://doi.org/10.1145/274644.274709}{Los Angeles, California, United
  States}, \href{https://doi.org/10.1145/274644.274709}{1998}.
  \href{https://doi.org/10.1145/274644.274709}
{doi: {{%
10\hspace{.1pt}\discretionary{.}{%
}{.}\hspace{.4pt}1145\discretionary{/}{%
}{/}274644\hspace{.1pt}\discretionary{.}{%
}{.}\hspace{.4pt}274709}}}


\bibitem{saket2015map}
B.~Saket, C.~Scheidegger, S.~G. Kobourov, and K.~B{\"o}rner.
\newblock Map-based visualizations increase recall accuracy of data.
\newblock In {\em Computer Graphics Forum}, vol.~34, pp. 441--450. Wiley Online
  Library, 2015.

\bibitem{shaffer_overview_2017}
\href{https://doi.org/10.3389/fpubh.2017.00258}{F.~Shaffer and J.~P. Ginsberg}.
\newblock \href{https://doi.org/10.3389/fpubh.2017.00258}{An {Overview} of
  {Heart} {Rate} {Variability} {Metrics} and {Norms}}.
\newblock \href{https://doi.org/10.3389/fpubh.2017.00258}{{\em Frontiers in
  Public Health}}, \href{https://doi.org/10.3389/fpubh.2017.00258}{5:258},
  \href{https://doi.org/10.3389/fpubh.2017.00258}{Sept. 2017}.
  \href{https://doi.org/10.3389/fpubh.2017.00258}
{doi: {{%
10\hspace{.1pt}\discretionary{.}{%
}{.}\hspace{.4pt}3389\discretionary{/}{%
}{/}fpubh\hspace{.1pt}\discretionary{.}{%
}{.}\hspace{.4pt}2017\hspace{.1pt}\discretionary{.}{%
}{.}\hspace{.4pt}00258}}}


\bibitem{sweller1988cognitive}
J.~Sweller.
\newblock Cognitive load during problem solving: Effects on learning.
\newblock {\em Cognitive science}, 12(2):257--285, 1988.

\bibitem{sweller1998cognitive}
J.~Sweller, J.~J. Van~Merrienboer, and F.~G. Paas.
\newblock Cognitive architecture and instructional design.
\newblock {\em Educational psychology review}, 10(3):251--296, 1998.

\bibitem{trujillo_theta_2007}
\href{https://doi.org/10.1016/j.clinph.2006.11.009}{L.~T. Trujillo and J.~J.~B.
  Allen}.
\newblock \href{https://doi.org/10.1016/j.clinph.2006.11.009}{Theta {EEG}
  dynamics of the error-related negativity}.
\newblock \href{https://doi.org/10.1016/j.clinph.2006.11.009}{{\em Clinical
  Neurophysiology: Official Journal of the International Federation of Clinical
  Neurophysiology}},
  \href{https://doi.org/10.1016/j.clinph.2006.11.009}{118(3):645--668},
  \href{https://doi.org/10.1016/j.clinph.2006.11.009}{Mar. 2007}.
  \href{https://doi.org/10.1016/j.clinph.2006.11.009}
{doi: {{%
10\hspace{.1pt}\discretionary{.}{%
}{.}\hspace{.4pt}1016\discretionary{/}{%
}{/}j\hspace{.1pt}\discretionary{.}{%
}{.}\hspace{.4pt}clinph\hspace{.1pt}\discretionary{.}{%
}{.}\hspace{.4pt}2006\hspace{.1pt}\discretionary{.}{%
}{.}\hspace{.4pt}11\hspace{.1pt}\discretionary{.}{%
}{.}\hspace{.4pt}009}}}


\bibitem{ware2005supporting}
C.~Ware and R.~Bobrow.
\newblock Supporting visual queries on medium-sized node--link diagrams.
\newblock {\em Information Visualization}, 4(1):49--58, 2005.

\bibitem{ware2002cognitive}
C.~Ware, H.~Purchase, L.~Colpoys, and M.~McGill.
\newblock Cognitive measurements of graph aesthetics.
\newblock {\em Information visualization}, 1(2):103--110, 2002.

\bibitem{werkle-bergner_cortical_2006}
\href{https://doi.org/10.1016/j.neubiorev.2006.06.009}{M.~Werkle-Bergner,
  V.~Müller, S.-C. Li, and U.~Lindenberger}.
\newblock \href{https://doi.org/10.1016/j.neubiorev.2006.06.009}{Cortical {EEG}
  correlates of successful memory encoding: implications for lifespan
  comparisons}.
\newblock \href{https://doi.org/10.1016/j.neubiorev.2006.06.009}{{\em
  Neuroscience and Biobehavioral Reviews}},
  \href{https://doi.org/10.1016/j.neubiorev.2006.06.009}{30(6):839--854},
  \href{https://doi.org/10.1016/j.neubiorev.2006.06.009}{2006}.
  \href{https://doi.org/10.1016/j.neubiorev.2006.06.009}
{doi: {{%
10\hspace{.1pt}\discretionary{.}{%
}{.}\hspace{.4pt}1016\discretionary{/}{%
}{/}j\hspace{.1pt}\discretionary{.}{%
}{.}\hspace{.4pt}neubiorev\hspace{.1pt}\discretionary{.}{%
}{.}\hspace{.4pt}2006\hspace{.1pt}\discretionary{.}{%
}{.}\hspace{.4pt}06\hspace{.1pt}\discretionary{.}{%
}{.}\hspace{.4pt}009}}}


\bibitem{white2012brain}
D.~J. White, M.~Congedo, J.~Ciorciari, and R.~B. Silberstein.
\newblock Brain oscillatory activity during spatial navigation: theta and gamma
  activity link medial temporal and parietal regions.
\newblock {\em Journal of cognitive neuroscience}, 24(3):686--697, 2012.

\bibitem{oursurvey}
\href{https://doi.org/https://doi.org/10.1016/j.visinf.2018.12.006}{V.~Yoghourdjian,
  D.~Archambault, S.~Diehl, T.~Dwyer, K.~Klein, H.~C. Purchase, and H.-Y. Wu}.
\newblock
  \href{https://doi.org/https://doi.org/10.1016/j.visinf.2018.12.006}{Exploring
  the limits of complexity: A survey of empirical studies on graph
  visualisation}.
\newblock
  \href{https://doi.org/https://doi.org/10.1016/j.visinf.2018.12.006}{{\em
  Visual Informatics}},
  \href{https://doi.org/https://doi.org/10.1016/j.visinf.2018.12.006}{2(4):264--282},
  \href{https://doi.org/https://doi.org/10.1016/j.visinf.2018.12.006}{2018}.
  \href{https://doi.org/10.1016/j.visinf.2018.12.006}
{doi: {{%
10\hspace{.1pt}\discretionary{.}{%
}{.}\hspace{.4pt}1016\discretionary{/}{%
}{/}j\hspace{.1pt}\discretionary{.}{%
}{.}\hspace{.4pt}visinf\hspace{.1pt}\discretionary{.}{%
}{.}\hspace{.4pt}2018\hspace{.1pt}\discretionary{.}{%
}{.}\hspace{.4pt}12\hspace{.1pt}\discretionary{.}{%
}{.}\hspace{.4pt}006}}}


\bibitem{yoghourdjian2018graph}
V.~Yoghourdjian, T.~Dwyer, K.~Klein, K.~Marriott, and M.~Wybrow.
\newblock Graph thumbnails: Identifying and comparing multiple graphs at a
  glance.
\newblock {\em IEEE Transactions on Visualization and Computer Graphics}, 2018.

\end{thebibliography}
\end{document}